\documentclass[11pt]{article}
\usepackage[T1]{fontenc}
\usepackage[utf8]{inputenc}
\usepackage{lmodern}
\usepackage[margin=1in]{geometry}
\usepackage{graphicx}
\usepackage{yfonts}
\usepackage{subcaption}
\usepackage{caption}
\usepackage{booktabs}
\usepackage{float}
\usepackage{bm}
\usepackage{amsmath,amssymb,amsthm,mathrsfs}
\usepackage[authoryear,round]{natbib}
\usepackage[colorlinks=true,linkcolor=blue,citecolor=blue,urlcolor=blue]{hyperref}

\def\half{{\textstyle\frac{1}{2}}}
\DeclareMathOperator{\rd}{d}
\newtheorem{proposition}{Proposition}

\title{Agreement coefficients for continuous variables: A review}
\author{Ronny O. Vallejos\\
Departamento de Matemática, Universidad Técnica Federico Santa María\\
Avenida Espa\~na 1680, Valparaíso, Chile\\
\texttt{ronny.vallejos@usm.cl}}
\date{}

\begin{document}
\maketitle

\maketitle


\section*{Abstract}
Agreement coefficients provide a fundamental framework for quantifying the concordance between two or more measurement methods applied to the same continuous variable. Unlike correlation, which measures the strength of a linear relationship, agreement focuses on assessing whether measurements are numerically similar, capturing both precision and accuracy. This review provides a comprehensive overview of the primary statistical approaches for assessing agreement between continuous variables. Such a synthesis is timely, as it has been 15–20 years since the last major review in the field. Beginning with the seminal contributions of Bland and Altman (1986) and Lin (1989), the paper discusses extensions of their methods to robust, multivariate, and repeated-measures settings, as well as recent developments like the probability of agreement and measures based on alternative distance functions measures. Special attention is given to probabilistic and spatial generalizations, including frameworks designed for geostatistical and areal data, which have become increasingly relevant in modern applications such as image analysis and environmental statistics. Through illustrative examples and comparative discussions, this review highlights the evolution, connections, and limitations of existing agreement measures, identifying open challenges and directions for future research.

{\it keywords:} Agreement coefficients, Concordance correlation coefficient, Probability of agreement, Spatial concordance.

\section{Introduction }\label{sec:intro}
Traditionally, when the primary objective is to evaluate the concordance between two measurement instruments, practitioners construct a $Y$ versus 
$X$ plot and quantify the deviation of the observed data points from the straight line passing through the origin. This graphical approach provides a simple yet effective means of examining the degree of agreement between two continuous variables in practice. Because visual interpretation of scatterplot patterns—such as those previously described—can be subjective across observers, efforts have been directed toward developing coefficients that quantify the degree of agreement, in a manner analogous to correlation.

Given the diverse applications of agreement measures across various contexts and data types, this study focuses on two continuous variables, which may increase in complexity when observations are georeferenced. 

Beyond the statistical techniques described in this paper, the need for further research is driven by the rapid evolution of data types. This evolution complicates the generalization of existing agreement coefficients and the definition of new distance measures for assessing concordance between objects. For instance, quantifying agreement between spatial networks remains a significant challenge that requires a meaningful definition of concordance involving more complex data structures.

\subsection{Historical Developments}\label{Hist}

We begin by emphasizing that agreement is not equivalent to linear correlation. Nevertheless, the Pearson correlation coefficient \citep{Pearson:1896} has historically been used to quantify the level of agreement between two continuous variables. Numerous studies in the literature illustrate that agreement refers to the degree of concordance between two variables, often conceptualized as the departure from the 45° line passing through the origin. In contrast, correlation assesses departure from any straight line, not necessarily one that intersects the origin, and is therefore a more general concept. On the other hand, the paired t-test \citep{Fisher:1925} has also been used to assess agreement; however, its results are highly dependent on the variances of the populations involved. In a similar vein, the test of zero correlation, with inference based on Fisher’s 
$z$-transformation \citep{Fisher:1915}, has also been applied in the context of agreement. Although there is nothing inherently incorrect about using such tests, the resulting conclusions do not align with the primary objectives of a method comparison study, in which the aim is often to compare two or more instruments designed to measure the same variable under similar conditions.

The articles by \cite{Bland:1986} and \cite{Lin:1989} are widely regarded as seminal references in the field of agreement measurement between two quantitative methods. The work of Bland and Altman is particularly known for introducing the limits of agreement and the accompanying plot of differences versus averages, now commonly referred to as the Bland–Altman plot. Although the authors had presented the methodology in an earlier publication, their 1986 paper popularized it within a medical context, and it remains the most frequently used technique in health and medical research.
Lin’s (1989) article is widely recognized for introducing the concordance correlation coefficient (CCC), which, while sharing certain properties with the Pearson correlation coefficient, provides a more appropriate measure of deviation from the 45° line passing through the origin. Since its introduction, the CCC has garnered considerable attention in the statistical literature, and numerous generalizations and adaptations for various contexts have been developed based on this foundational work.

The statistical literature on this topic has expanded rapidly since the publication of these two seminal papers. Several comprehensive reviews are available \citep{Barnhart:2007, Lin:2008, Choudhary:2009}, and the subject is also covered extensively in a number of books \citep{Dunn:2004, Broemeling:2009, Shoukri:2010, Lin:2012, Choudhary:2017}. Extensions of this type of agreement technique are varied and widely dispersed throughout the literature. Below, we highlight several generalizations of Lin’s concordance coefficient that have been developed primarily for a statistical audience.

\cite{King:2001} developed a robust version of Lin’s concordance coefficient based on alternative distance functions that are less sensitive to outliers and allow for the consideration of variants that may perform better under different scenarios. Later, \cite{Tashakor:2019} focused on functionals that yield robust 
$L$-statistics and provided real-data applications illustrating the practical performance of the $L_1$ measure. More recently, \cite{Vallejos:2025b} introduced a new version of an $L_1$	
  agreement coefficient that is robust and does not depend on tunning parameters, thereby simplifying its implementation. \cite{Bulut:2026} suggested a robust reformulation of the CCC using minimum covariance determinant estimators, preserving interpretability while improving robustness to outliers, as demonstrated through simulation studies and real-data applications.

\cite{King:2007} introduced a repeated-measures concordance coefficient that not only satisfies the essential properties for quantifying overall agreement between two $n\times 1$ vectors of random variables but also provides a more intuitive interpretation than the method previously proposed by \cite{Chinchilli:1996}. Subsequently, \cite{Hiriote:2011} extended Lin’s concordance coefficient to accommodate multivariate observations in the context of repeated measures. Further developments include the work of \cite{Leal:2019}, who examined a Lin-type coefficient for the assessment of local influence. 

An alternative framework for evaluating agreement between two sets of measurements is the probability of agreement (PA), first introduced in a series of papers by \cite{Stevens:2017}, \cite{Stevens:2018}, and \cite{Stevens:2020}. \cite{deCastro:2021} further advanced this line of research by developing Bayesian PA methods for comparing measurement systems under both homoscedastic and heteroscedastic error structures. The literature on PA is now substantial, with numerous methodological contributions and applications across diverse domains.

Extensions of agreement coefficients to spatial data have attracted growing attention in recent years. \cite{Vallejos:2020} proposed a framework for evaluating agreement between two continuous spatial processes when both variables are georeferenced. Under an increasing-domain sampling scheme, they established the asymptotic normality of the sample spatial concordance coefficient for a bivariate Gaussian process with a Wendland covariance function, and illustrated the method’s utility and limitations through an image analysis example. More recently, \cite{Acosta:2024} investigated a geostatistical generalization of the PA approach. Concordance indices have also been employed in neuroimaging to compare brain maps \citep{Alex:2018, Markello:2021}. Despite these advances, extending agreement indices to lattice data—where observations are linked to areal units rather than point locations has been addressed recently by \cite{Vallejos:2025}  motivated by the analysis of different methodologies for measuring poverty rates in Chile. 

\subsection{A Motivating Example}\label{motivating}

\begin{figure}[htbp]
    \centering

    \begin{subfigure}[b]{0.32\textwidth}
        \centering
        \includegraphics[width=\linewidth]{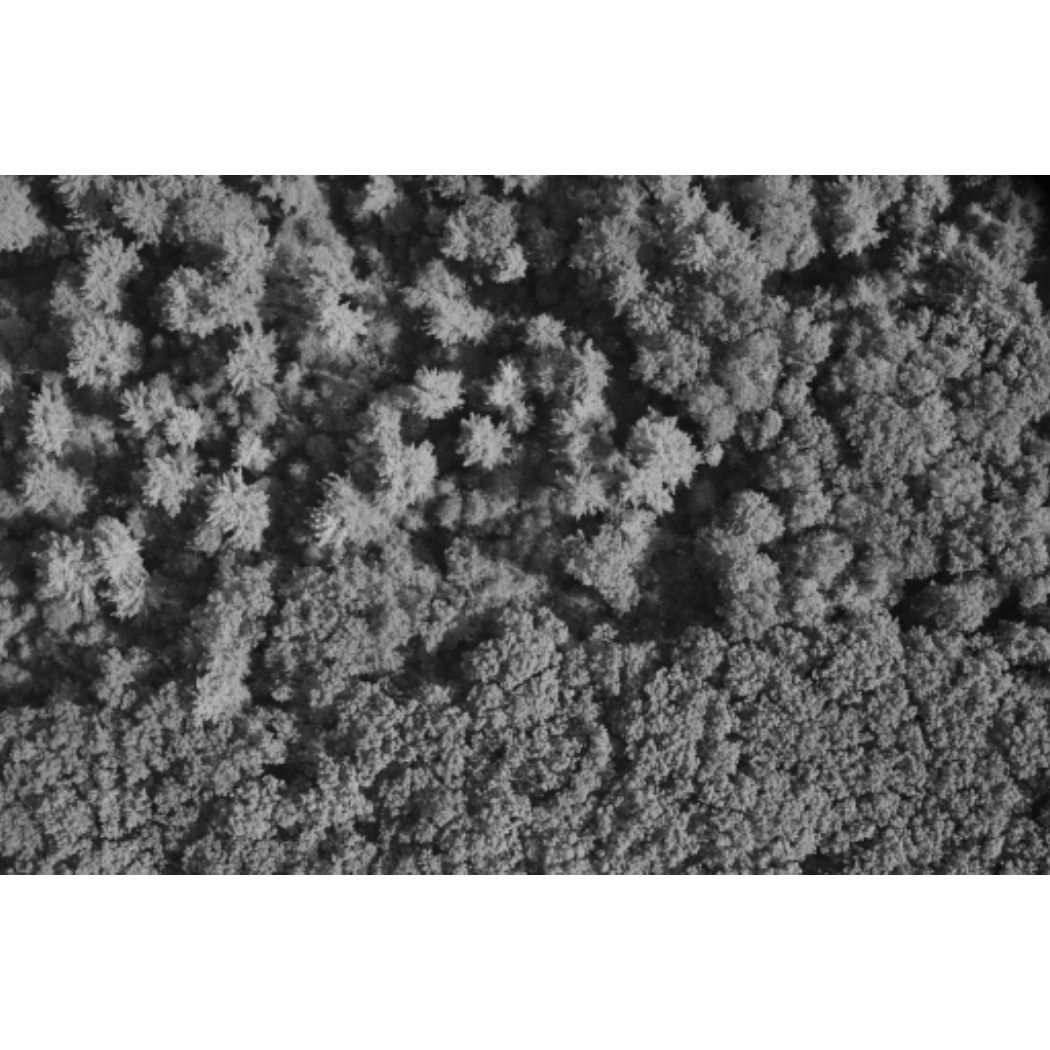}
        \caption{}
        \label{fig:forest-a}
    \end{subfigure}
    \hfill
    \begin{subfigure}[b]{0.38\textwidth}
        \centering
        \includegraphics[width=\linewidth]{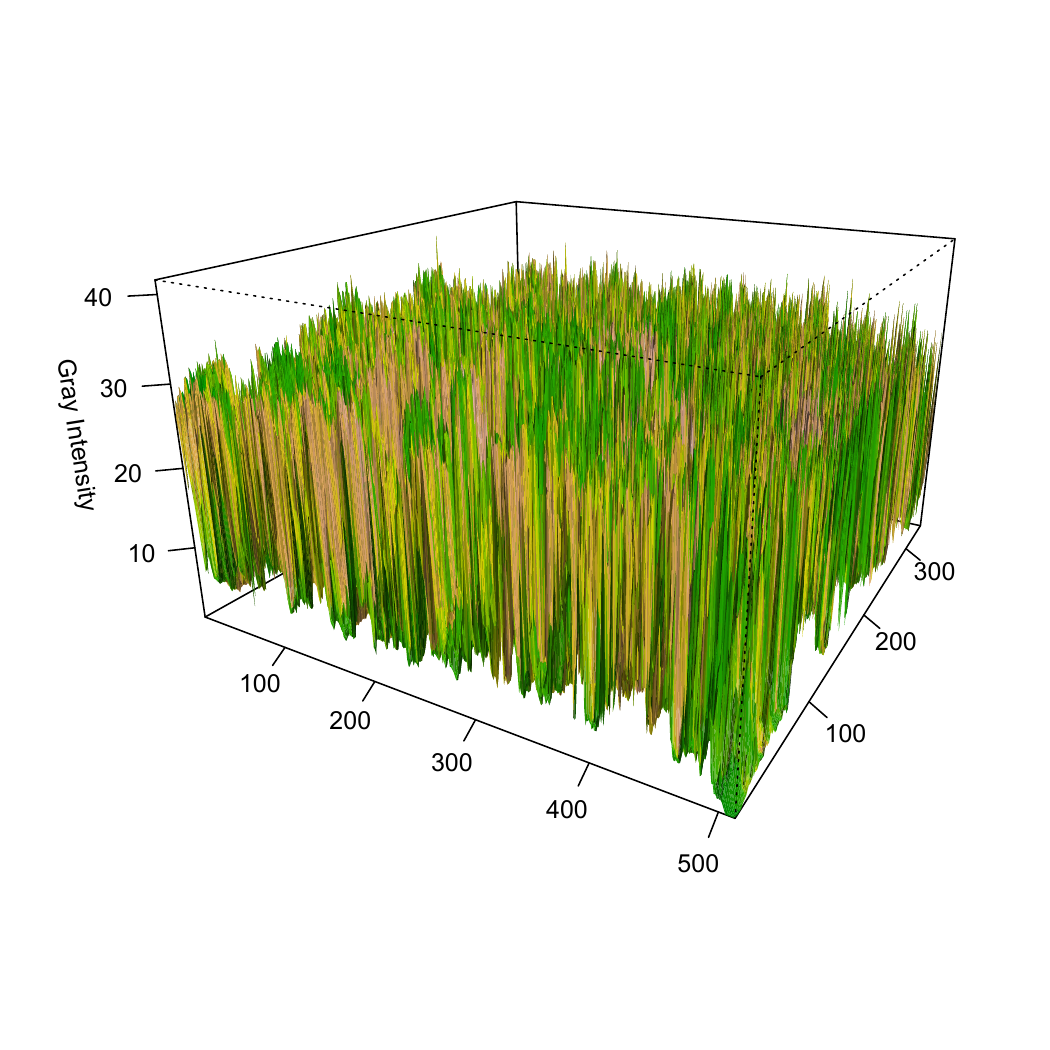}
        \caption{}
        \label{fig:forest-b}
    \end{subfigure}

    \vspace{0.6em}

    \begin{subfigure}[b]{0.235\textwidth}
        \centering
        \includegraphics[width=\linewidth]{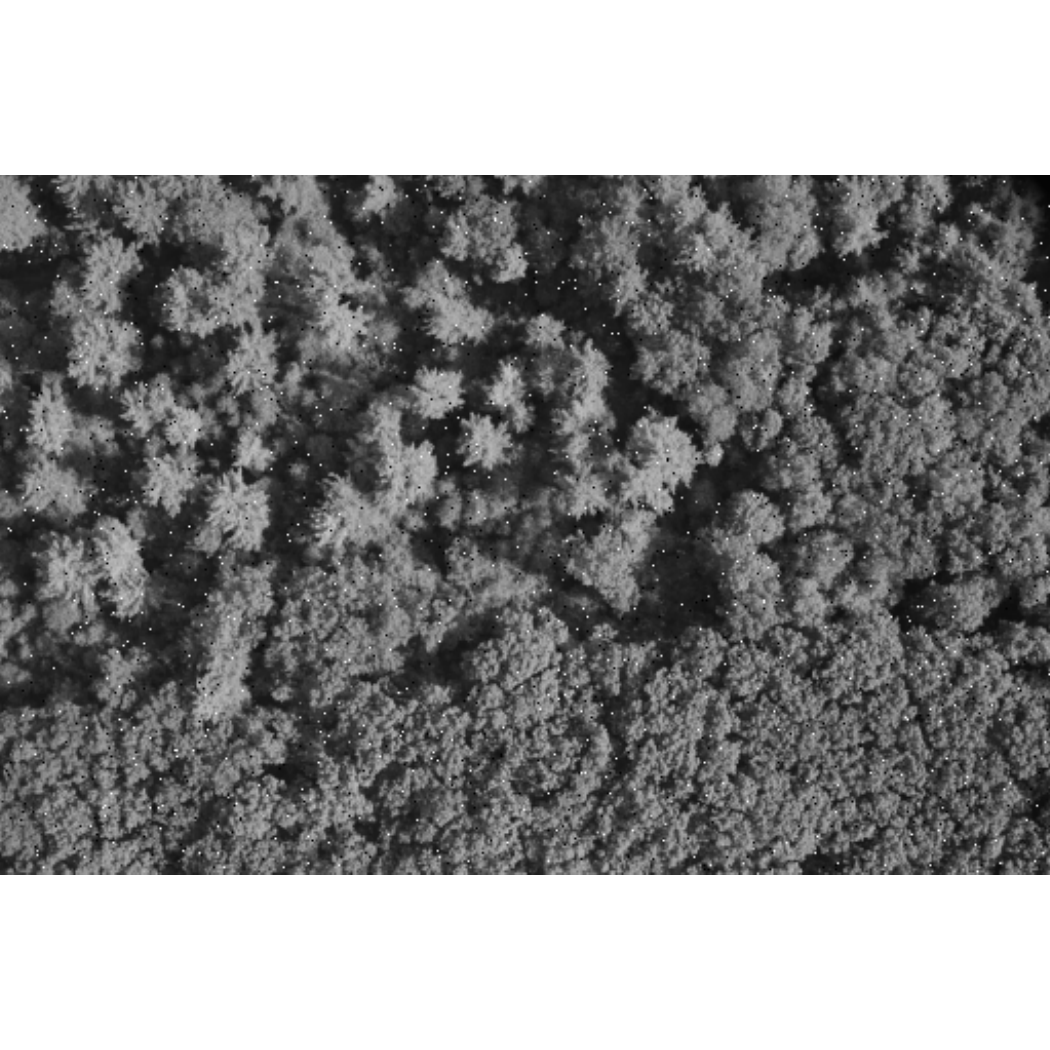}
        \caption{}
        \label{fig:forest-c}
    \end{subfigure}
    \hfill
    \begin{subfigure}[b]{0.235\textwidth}
        \centering
        \includegraphics[width=\linewidth]{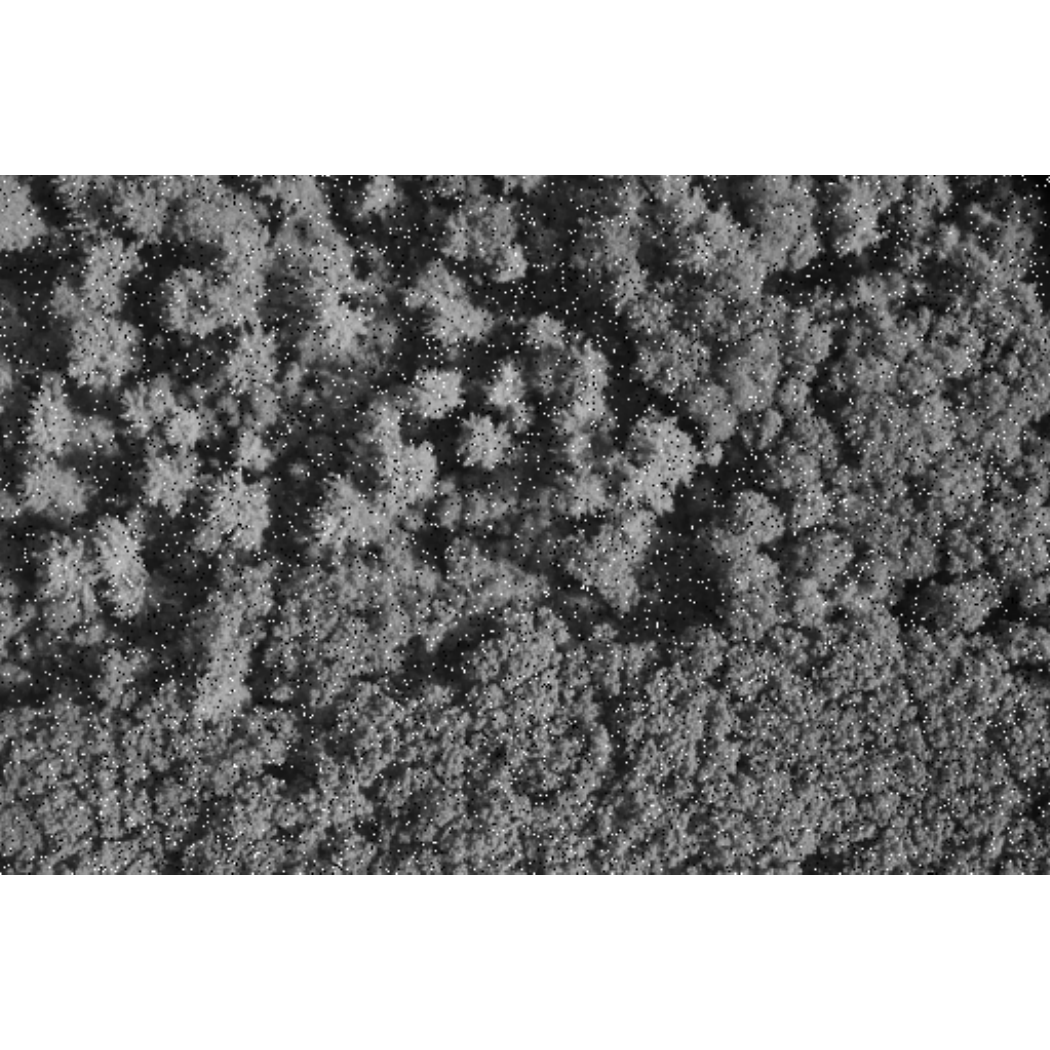}
        \caption{}
        \label{fig:forest-d}
    \end{subfigure}
    \hfill
    \begin{subfigure}[b]{0.235\textwidth}
        \centering
        \includegraphics[width=\linewidth]{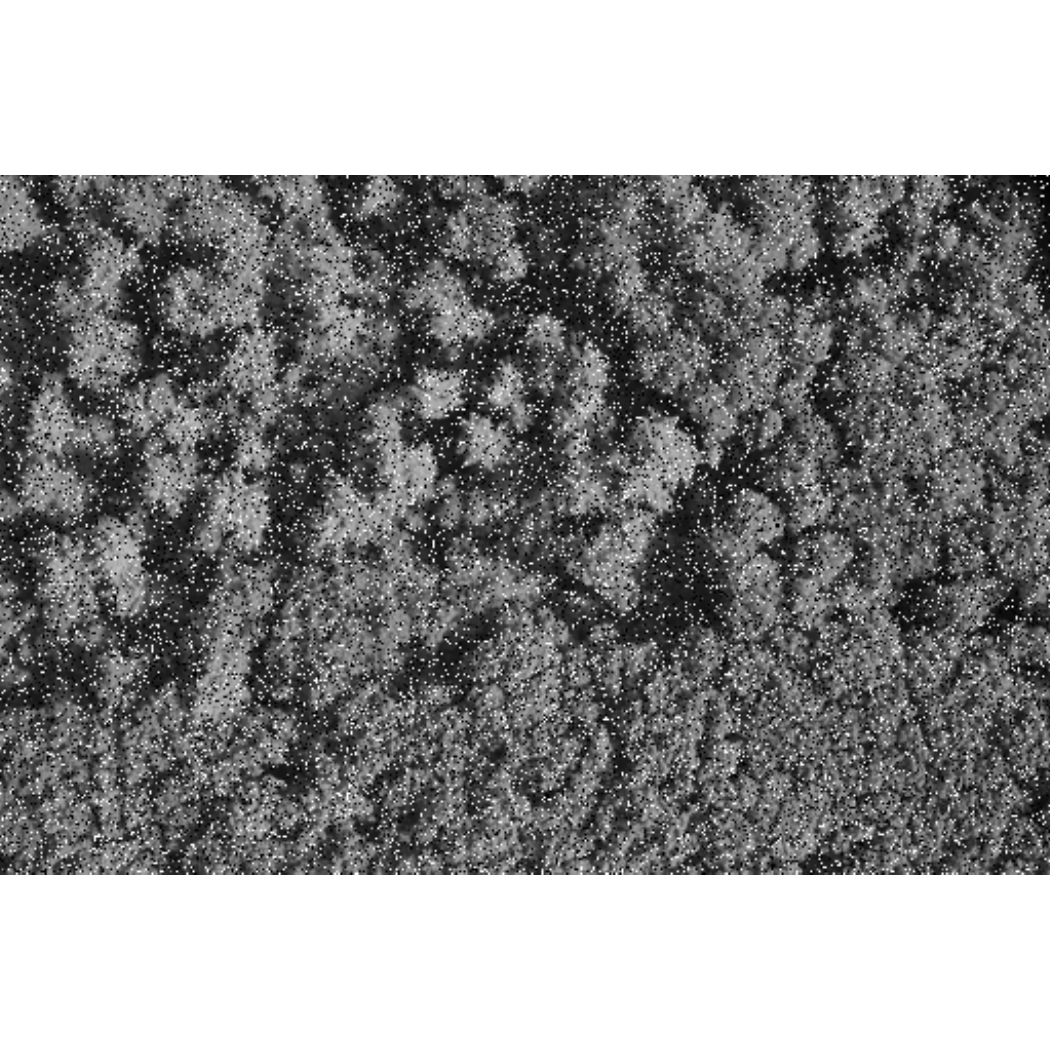}
        \caption{}
        \label{fig:forest-e}
    \end{subfigure}
    \hfill
    \begin{subfigure}[b]{0.235\textwidth}
        \centering
        \includegraphics[width=\linewidth]{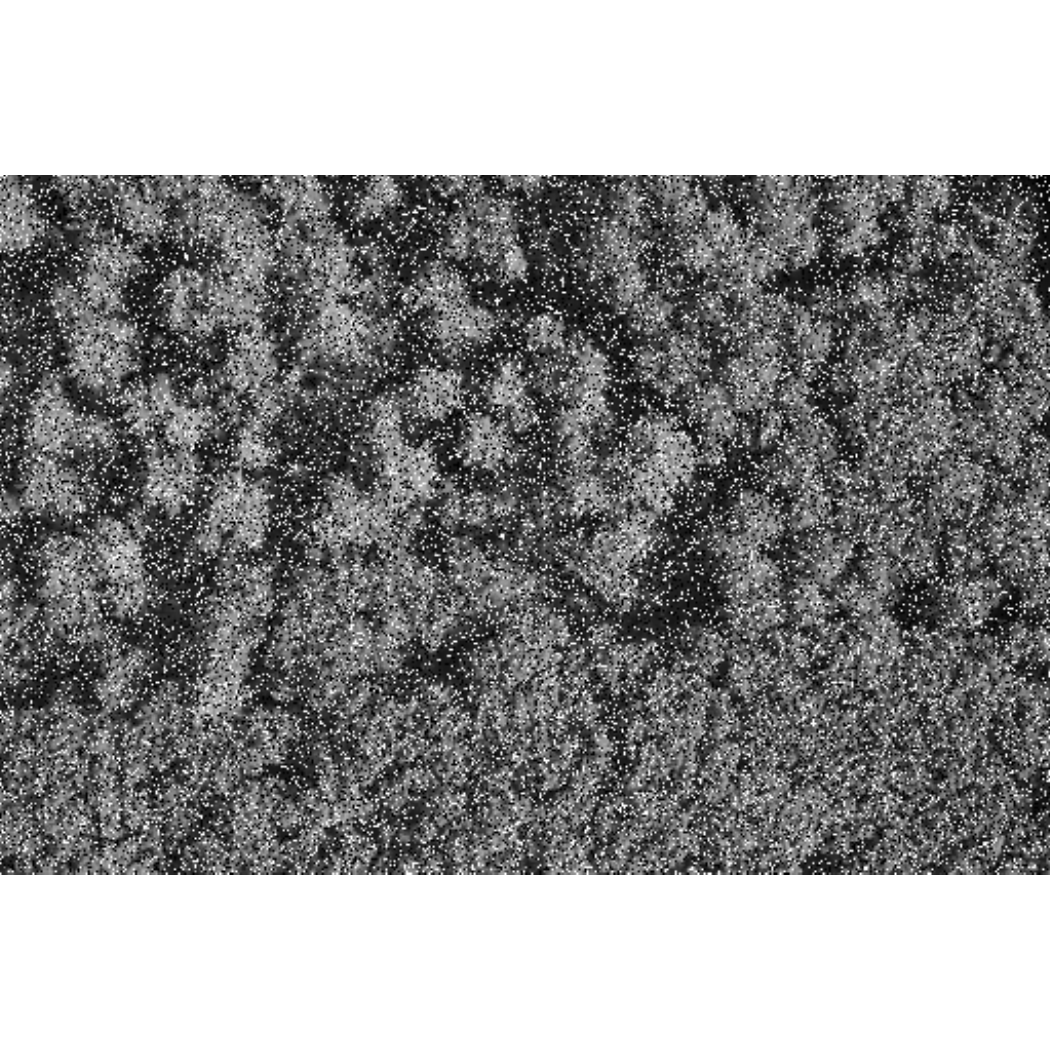}
        \caption{}
        \label{fig:forest-f}
    \end{subfigure}

    \vspace{0.6em}

    \begin{subfigure}[b]{0.235\textwidth}
        \centering
        \includegraphics[width=\linewidth]{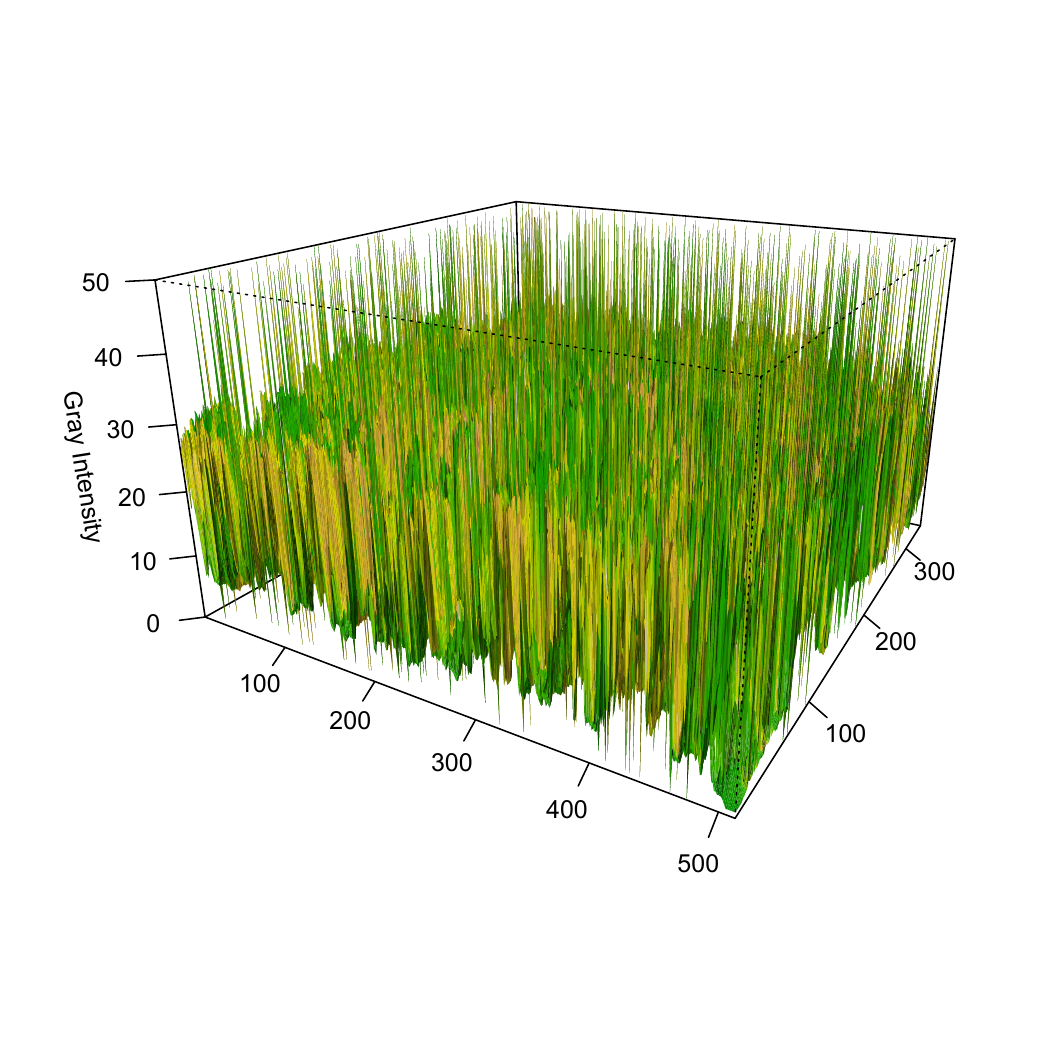}
        \caption{}
        \label{fig:forest-g}
    \end{subfigure}
    \hfill
    \begin{subfigure}[b]{0.235\textwidth}
        \centering
        \includegraphics[width=\linewidth]{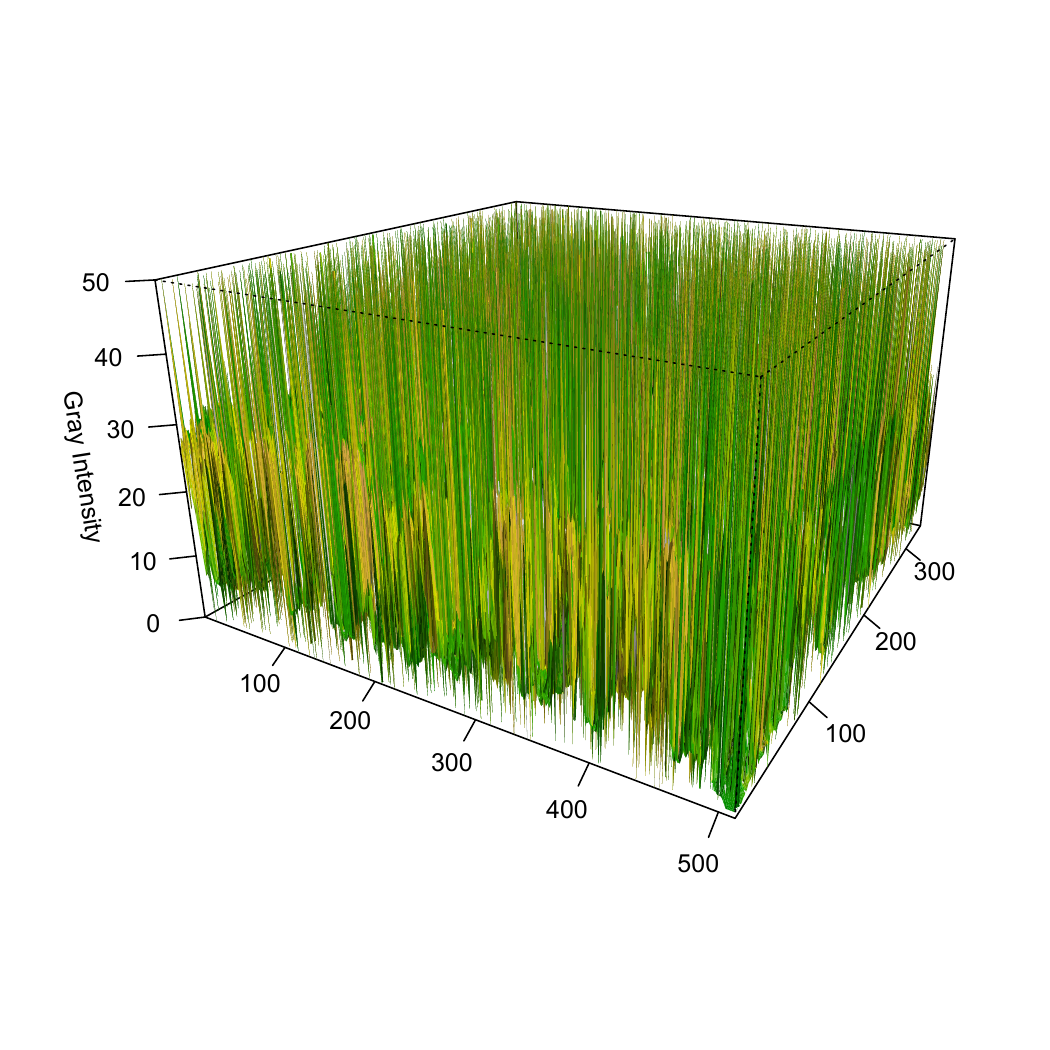}
        \caption{}
        \label{fig:forest-h}
    \end{subfigure}
    \hfill
    \begin{subfigure}[b]{0.235\textwidth}
        \centering
        \includegraphics[width=\linewidth]{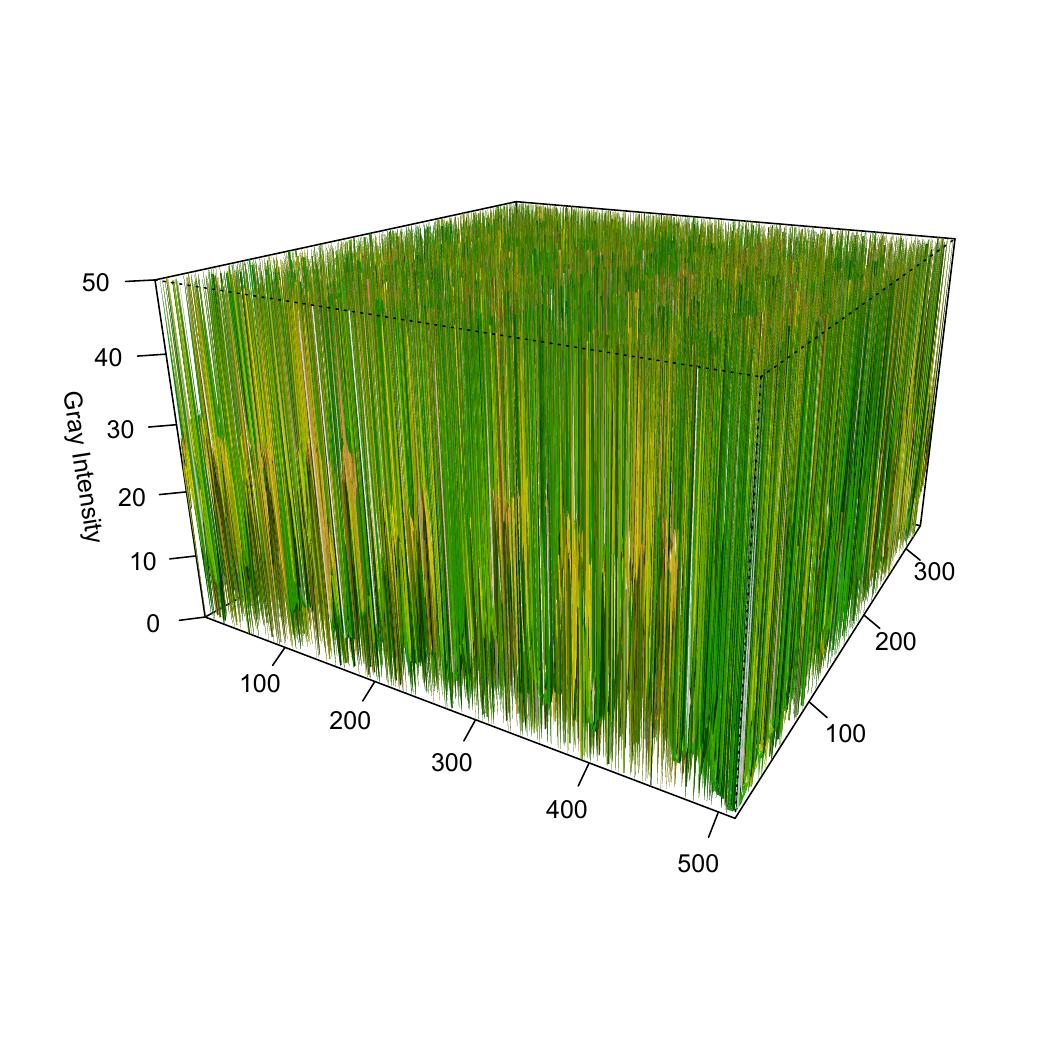}
        \caption{}
        \label{fig:forest-i}
    \end{subfigure}
    \hfill
    \begin{subfigure}[b]{0.235\textwidth}
        \centering
        \includegraphics[width=\linewidth]{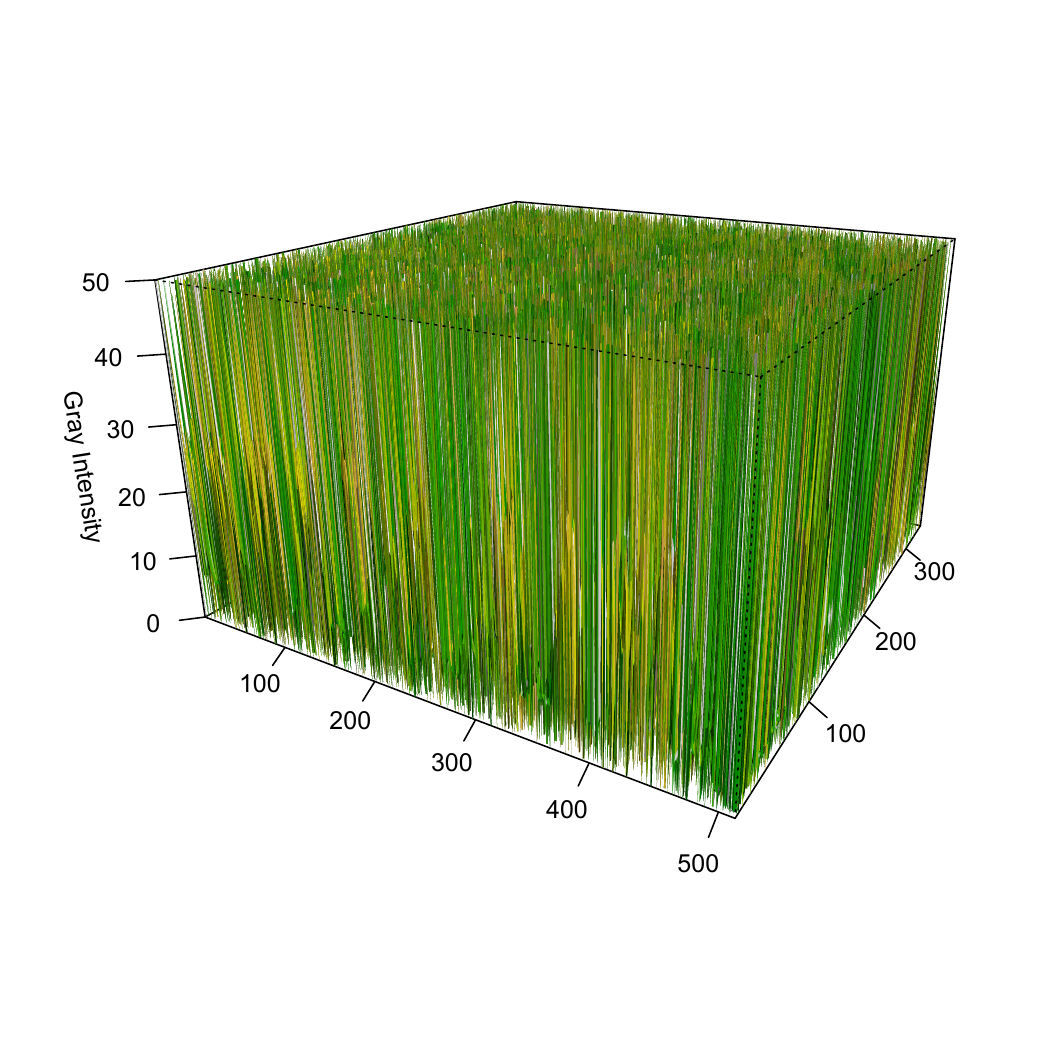}
        \caption{}
        \label{fig:forest-j}
    \end{subfigure}

    \caption{Panel of figures showing the visual effect of contamination of forest image. (a) Reference image taken above a section of forest at the Harvard Forest, Petersham, MA, USA and (b) its corresponding 3D perspective plot. (c-f) are the same image distorted with increasing amounts of salt-and-pepper noise. (g-j) show the change in gray intensity after the addition of salt-and-pepper noise to the images.}
    \label{fig:forest}
\end{figure}

Here, we present an illustrative example highlighting the differences among agreement, correlation, and similarity in images. To this end, we consider an aerial photograph of a forest stand in Petersham, MA. Figure~\ref{fig:forest}(a) displays a grayscale reference image with dimensions $341 \times 512$ pixels. This image was contaminated with salt-and-pepper noise \citep{Huang:2010}. Assume that $X$ denotes an image and that $X$ follows a zero-mean normal distribution with variance $\sigma^2$.
Then we consider additive noise following a zero-mean normal distribution with variance $\tau^2$ such that $\tau^2 \gg \sigma^2$. The contamination is located randomly in space such that a small percentage of observations are corrupted with probability $\delta$. 
Specifically,
\begin{equation}\label{eq:salt}
X \sim (1 - \delta)\, \mathcal{N}(0, \sigma^2) + \delta\, \mathcal{N}(0, \tau^2).
\end{equation}
The contamination scheme was generated assuming $\sigma^2 = 1$, $\tau^2 = 10$, and contamination proportions of $\delta = 0.01, 0.05, 0.15,$ and $0.25$. 
Figures~\ref{fig:forest}(c)--(f) display the contaminated versions of the original image corresponding to contamination levels of 1\%, 5\%, 15\%, and 25\%, respectively. In addition, Figures~\ref{fig:forest}(g)–(j) display perspective plots of the gray-level intensities of the contaminated images. As the level of contamination increases, the dispersion also increases, as evidenced by the spread along the 
$z$-axis in the three-dimensional scatter plots.

To gain further insight into the performance of the Pearson coefficient, the CCC, and the similarity index (SSIM) introduced by \citet{Wang:2004} (see Appendix \ref{app1}), these measures were computed for the images shown in Figure \ref{fig:forest}.

\begin{figure}[H]
    \centering
    \includegraphics[width=0.5\linewidth]{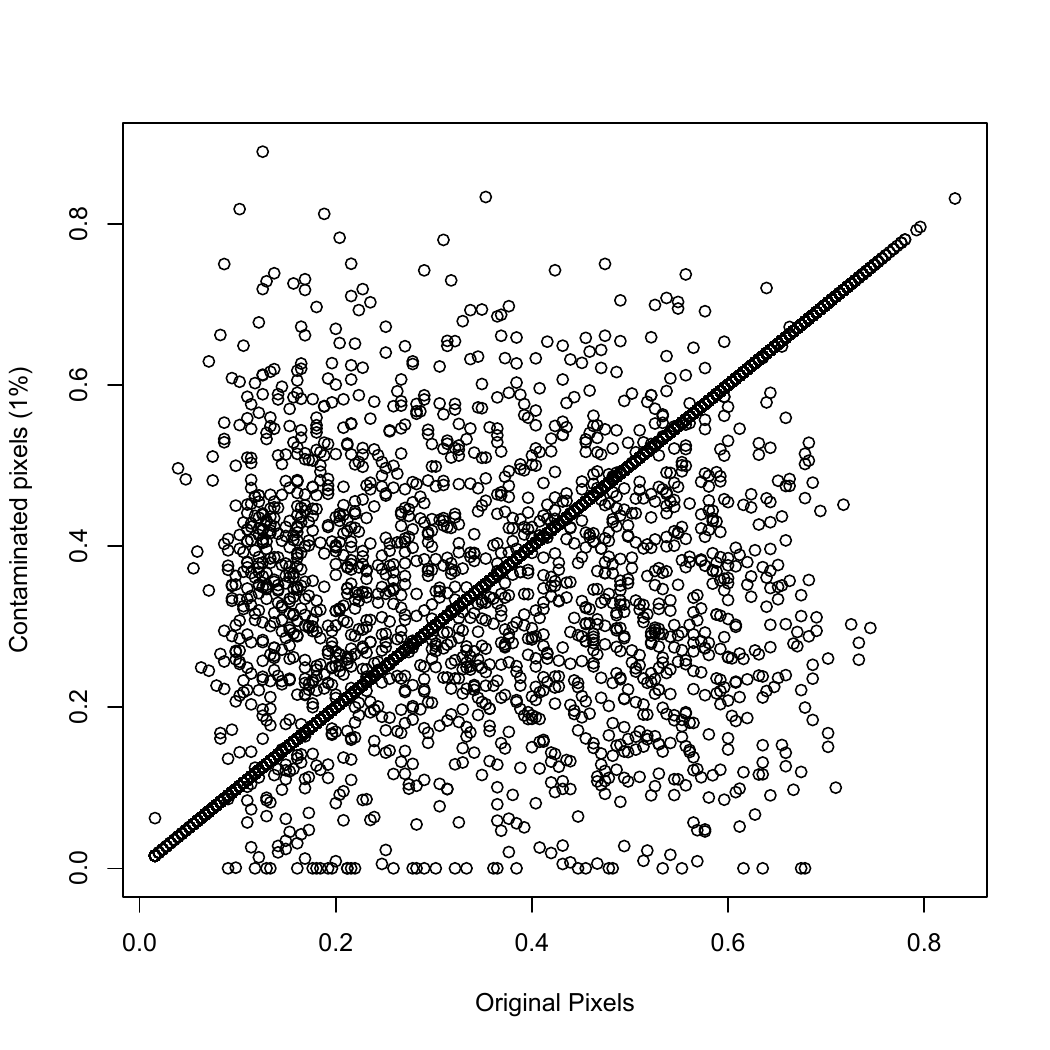}
    \caption{Effect of 1\% contamination on the gray label pixels of image displayed in Figure \ref{fig:forest}(a).}
    \label{fig:scatterplot}
\end{figure}

In Table \ref{tab:indices}, the sample values of the coefficients are summarized for different percentages of image contamination. Three main patterns can be observed. First, the Pearson correlation decreases as the level of contamination increases. Second, the CCC remains close to zero regardless of the contamination rate. Third, the SSIM index also decreases with increasing contamination, following a trend similar to that of the Pearson coefficient. Overall, these three indices capture different aspects of the relationship between the two images. As is well known, the Pearson coefficient quantifies linear association, the CCC assesses departures from the 45° line, and the SSIM evaluates similarity by jointly comparing means, variances, and correlation structures.
\begin{table}[h]
    \centering   
\begin{tabular}{lcccc} \toprule
Original & 1\% & 5\% & 15\% & 25\%\\ \midrule
Correlation & 0.959 & 0.812 & 0.597 & 0.458 \\ 
CCC   &-0.001 & -0.006& -0.019 & -0.006 \\
SSIM &0.959 & 0.812&0.565 &0.410\\
\bottomrule

\end{tabular} 
\caption{The Pearson correlation coefficient, Lin’s CCC, and the SSIM index computed for the images shown in Figure~\ref{fig:forest}.}
    \label{tab:indices}
\end{table}
To illustrate the effect of contamination on the CCC index, Figure \ref{fig:scatterplot} presents a scatterplot comparing the 1\% of contaminated pixels against the original image pixels.
Even a 1\% contamination, which produces a noticeable spread around the 45$^\circ$ line through the origin, is enough to drive the CCC close to zero.

\subsection{Layout of the Paper}\label{Layout}

The remainder of the paper is organized as follows. Section \ref{sec:cont} reviews three well-known concordance measures: the Bland–Altman limits of agreement, Lin’s concordance correlation coefficient, and the probability of agreement, along with extensions and connections to related techniques. Section \ref{sec:time} introduces the notion of agreement for pairs of time series and discusses its relationship with comovement in chronological data. Section \ref{sec:mult} describes the proposed agreement coefficients for multivariate random vectors. Agreement measures for spatial data are examined in Section \ref{sec:spatial}. An application in the context of air quality monitoring is presented in Section \ref{sec:application}. Finally, the paper concludes by outlining open problems and directions for future research. The proofs of two results established in Section \ref{sec:lin} are deferred to Appendix \ref{app2}.

\section{Statistical Agreement for Continuous Outcomes}
\label{sec:cont}

\subsection{Measuring the discrepancy between two random variables}
When comparing two methods or procedures, the underlying idea is to assess the relationship between two random variables, which can be approached from several perspectives. If \( X \) and \( Y \) denote the random variables under comparison, the discrepancy between them can be expressed as a function \( g \) such that
\[
D = g(X - Y).
\]
Most coefficients designed to measure correlation or agreement are based on this principle. For example, the Pearson correlation coefficient is derived from the inner product \( g(X, Y) = \mathbb{E}[XY] \), while Lin’s concordance correlation coefficient is based on \( g(X, Y) = \mathbb{E}[(X - Y)^2] \).
The choice of the function \( g \) in practice depends on several factors, including the robustness of the resulting coefficient, its theoretical properties, and the computational feasibility of its estimation.

\subsection{The Bland--Altman procedure}
The Bland–Altman method \citep{Bland:1986} is one of the most widely used techniques for quantifying agreement between two measurement methods. It provides an intuitive and informative assessment by directly quantifying and visualizing the extent of disagreement between them.

Consider paired observations $(X_i, Y_i)$, $i = 1, \ldots, n$, obtained from two different instruments or procedures applied to the same subjects. The analysis begins by computing, for each pair, the average value  
\[
\overline{M}_i = \frac{X_i + Y_i}{2},
\]
and the corresponding difference  
$D_i = X_i - Y_i$.
The mean of the differences  
\[
\overline{D} = \frac{1}{n} \sum_{i=1}^{n} D_i,
\]
provides an estimate of the systematic bias between the two methods, while the standard deviation of the differences,  
\[
s_D = \sqrt{\frac{1}{n - 1}\sum_{i=1}^{n}(D_i - \overline{D})^2},
\]
quantifies the random variation in their disagreement.  
Assuming that the differences $D_i$ follow an approximately normal distribution, Bland and Altman proposed the 95\% limits of agreement (LoA), defined as  
\[
\overline{D} \pm 1.96\, s_D,
\]
which represent the interval within which approximately 95\% of the differences between the two measurement methods are expected to lie. A small bias and narrow limits of agreement indicate good concordance between the methods.  

Graphically, the results are summarized through the Bland--Altman plot, which displays the differences $D_i$ on the vertical axis against the averages $\overline{M}_i$ on the horizontal axis. Horizontal reference lines corresponding to the mean difference ($\overline{D}$) and to the upper and lower limits of agreement ($\overline{D} \pm 1.96\, s_D$) are superimposed. This visualization facilitates the identification of systematic trends, such as proportional bias (when the differences depend on the magnitude of the measurements) or heteroscedasticity (when the variability of the differences changes with the measurement level).  

For illustration purposes, the 95\% Bland–Altman confidence interval was computed, and the corresponding Bland–Altman plot was generated for the images shown in Figures \ref{fig:forest}(a) and (c), with the latter representing 1\% contamination.

\begin{figure}
    \centering
    \includegraphics[width=0.5\linewidth]{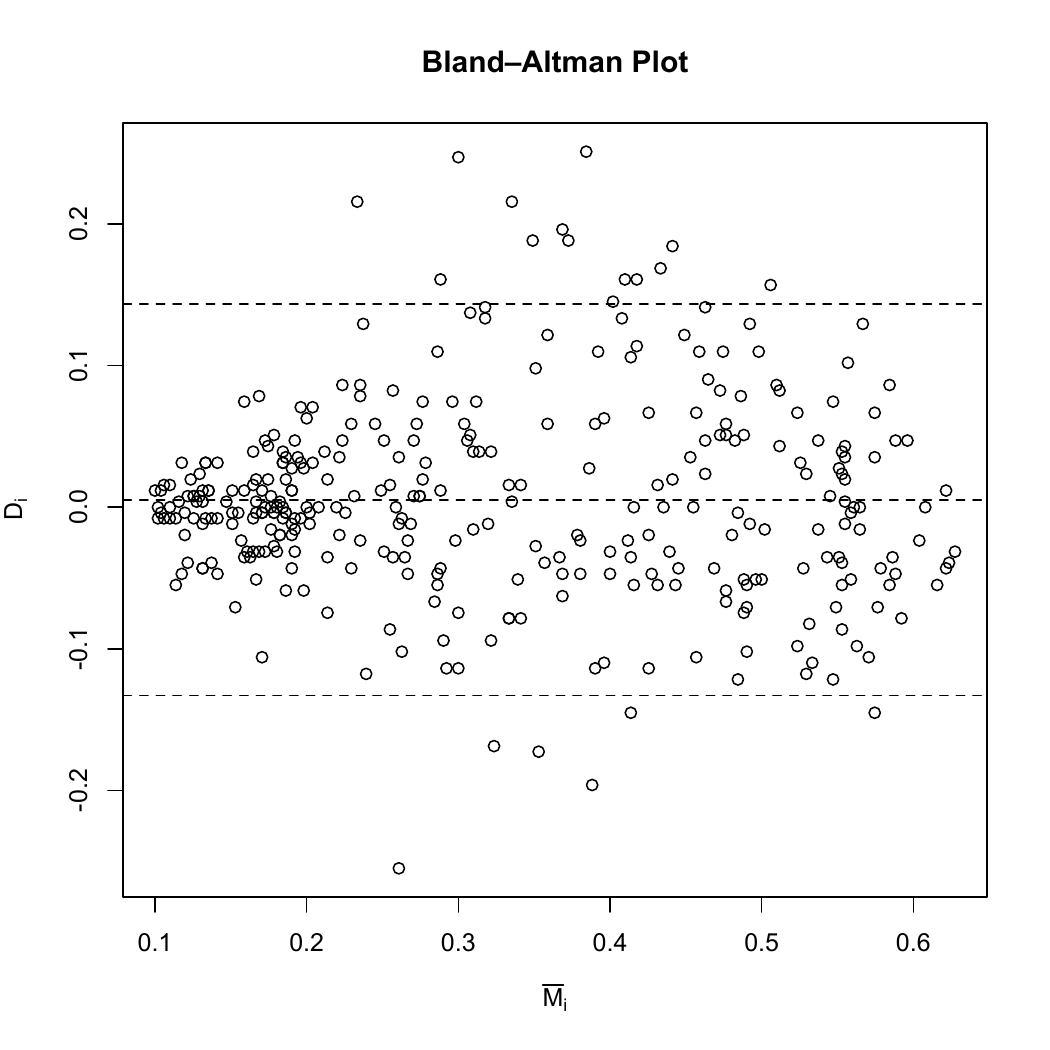}
    \caption{Bland–Altman plot comparing the images displayed in Figures \ref{fig:forest}(a) and (c). The dotted lines indicate the 95\% LoA.}
    \label{fig:BA_plot}
\end{figure}

Figure \ref{fig:BA_plot} displays a considerable dispersion of the $D_i$ values, with several points lying outside the confidence interval. While the effect of contamination is evident, identifying the specific impact of outliers is more complex, as both $\overline{M}_i$ and $D_i$ may have been influenced by extreme values. A more robust approach would involve replacing the classical estimators of $\overline{M}_i$ and $\overline{D}$ with robust counterparts, leading to a corresponding robust version of the confidence intervals. On the other hand, the images used to measure agreement may exhibit spatial correlation, which has not been considered in the analysis. This issue will be addressed in Section \ref{sec:spatial}.

Although limits of agreement are a standard and well-established tool for assessing agreement, the topic remains an active area of research. For instance, \cite{Taffe:2025} introduced a Stata package specifically designed to evaluate agreement between two continuous measurement methods using clinical tolerance limits—user-defined bounds that represent clinically acceptable differences.
 
\subsection{Lin's Coefficient and its generalizations}
\label{sec:lin}

The CCC, proposed by \citet{Lin:1989}, is a statistical measure designed to quantify the agreement between two continuous variables by jointly assessing both precision and accuracy. Precision reflects how closely the data points cluster around the line of perfect agreement, whereas accuracy is assessed through a correction factor that accounts for systematic bias.
Let $(X_i, Y_i)$, $i = 1, \ldots, n$, denote paired observations obtained from two measurement methods applied to the same subjects. Denote the sample means, variances, and covariance as  
\[
\overline{X} = \frac{1}{n}\sum_{i=1}^n X_i, \qquad \overline{Y} = \frac{1}{n}\sum_{i=1}^n Y_i,
\]
\[
S_X^2 = \frac{1}{n-1}\sum_{i=1}^n (X_i - \overline{X})^2, \qquad S_Y^2 = \frac{1}{n-1}\sum_{i=1}^n (Y_i -\overline{Y} )^2, \qquad S_{XY} = \frac{1}{n-1}\sum_{i=1}^n (X_i - \overline{X})(Y_i - \overline{Y}).
\]
The sample CCC is defined as  
\[
\widehat{\rho}_c = \frac{2S_{XY}}{S_X^2 + S_Y^2 + (\overline{X} - \overline{Y})^2}.
\]
The population counterpart when the random vector $(X,Y)^{\top}\sim \mathcal{N}(\bm \mu,\Sigma)$ where $\bm \mu=(\mu_X,\mu_Y)^{\top}$ and \begin{equation}\label{Sigma}
\bm \Sigma=\begin{pmatrix} \sigma_{X}^2 & \sigma_{XY}\\ \sigma_{YX} & \sigma_{Y}^2\end{pmatrix},\end{equation}
is defined as
\begin{equation*}
    \rho_c=1-\frac{\mathbb{E}[(X-Y)^2]}{\mathbb{E}[(X-Y)^2\mid\sigma_{XY}=0]}=\frac{2 \sigma_{XY}}{\sigma_X^2+\sigma_Y^2+\left(\mu_X-\mu_Y\right)^2}.
\end{equation*}
 The coefficient $\rho_c$ takes values in $[-1, 1]$, where $\rho_c = 1$ indicates perfect concordance (all points lying on the 45$^\circ$ line), $\rho_c = 0$ denotes absence of concordance, and $\rho_c = -1$ corresponds to perfect discordance. The CCC can also be decomposed as 
\[
\rho_c = \rho \, C_b,
\]
where $\rho = \sigma_{XY} / (\sigma_X \sigma_Y)$ is the Pearson correlation coefficient, measuring precision,   
$C_b =\{(v+1/v+u^2)/2\}^{-1}$, $v=\sigma_X/\sigma_Y$ and $u=(\mu_X-\mu_Y)/\sqrt{\sigma_X\sigma_Y}$. Here $C_b$
is the bias correction factor that quantifies accuracy by penalizing deviations from the 45$^\circ$ line of perfect concordance. Thus, high concordance requires both $\rho \approx 1$ and $C_b \approx 1$.  

Under the assumption that $(X_i, Y_i)^{\top}$, $i=1,\ldots,n$, are independent and identically distributed according to a bivariate normal distribution $\mathcal{N}(\boldsymbol{\mu}, \boldsymbol{\Sigma})$, where $\boldsymbol{\mu}=(\mu_X,\mu_Y)^{\top}$ and $\boldsymbol{\Sigma}$ is given in \eqref{Sigma}, the estimator $\widehat{\rho}_c$ is consistent for $\rho_c$ and asymptotically normal.

Inference for this coefficient can be conducted using Fisher’s transformation. Specifically, \cite{Lin:1989} showed that
\[
\widehat{Z}
=\frac{1}{2}\log\!\left(\frac{1+\widehat{\rho}_c}{1-\widehat{\rho}_c}\right)
\xrightarrow{\mathcal{D}} \mathcal{N}(Z,\sigma_Z^2),
\qquad \text{as } n\to\infty,
\]
where
\[
Z=\tanh^{-1}(\rho_c)
=\frac{1}{2}\log\!\left(\frac{1+\rho_c}{1-\rho_c}\right),
\]
and
\[
\sigma_{\widehat{Z}}^2
=\frac{1}{n-2}\left[
\frac{(1-\rho^2)\rho_c^2}{(1-\rho_c^2)\rho^2}
+\frac{4v^2(1-\rho_c)\rho_c^3}{(1-\rho_c^2)^2\rho}
+\frac{2v^4\rho_c^4}{(1-\rho_c^2)^2\rho^2}
\right].
\]

As a consequence of the asymptotic normality of the sample CCC, an approximate hypothesis testing problem of the form $$\text{H}_0:\rho_c=\rho_0 ~ \text{versus} ~ \text{H}_1:\rho_c \neq \rho_0$$ for a fixed $\rho_0$ can be constructed. Alternatively, an approximate confidence interval for $\rho_c$  of the form
$$[\tanh(Z-z_{\alpha/2}\sigma_{\widehat{Z}}),\tanh(Z+z_{\alpha/2}\sigma_{\widehat{Z}})]$$
can be used, where $z_{\alpha/2}$ is the upper quantile of order $\alpha/2$ of the standard normal distribution.

The asymptotic variance tends to increase as the true concordance decreases, implying that the precision of the CCC estimate diminishes for lower agreement levels. Consequently, while the CCC provides a unified measure of both correlation and bias, care must be taken when interpreting its confidence intervals, especially in the presence of heteroscedasticity or non-normal measurement errors. A detailed discussion of agreement measures in diverse contexts can be found in \cite{Lin:2002}. 

One method for evaluating the impact of preprocessing—such as applying a Box–Cox transformation to symmetrize a dataset—is to consider a general transformation of \(X\). Precisely, assume that $Y=g(X)$, where $g(\cdot)$ is a differentiable function. Considering a first order Taylor expansion of $g(\cdot)$  about $\mu_1$, we have that
\begin{equation}\label{eq:trans1}
\rho_c(X,g(X)) \approx 
\frac{2\,g'(\mu_1)\,\sigma_X^{2}}
{\sigma_X^{2}\!\left[1+\big(g'(\mu_X)\big)^{2}\right]
+\left(\mu_X - g(\mu_X)\right)^{2}}.
\end{equation}
This expression clarifies how a transformation affects concordance, and provides a quick approximation without requiring numerical recomputation of the CCC.
In the absence of a local shift, i.e., $g(\mu_X)=\mu_X$ concordance is determined entirely by the local slope of the transformation and can be expressed as
$$
\rho_c(X,g(X)) \approx \frac{1 + \big(g'(\mu_X)\big)^2}{2\,g'(\mu_X)}.
$$
Now, consider a nearly linear transformation of the form $g(x)=x+\epsilon h(x),$ where $\epsilon<<1$ and $h(\cdot)$ is a differentiable function. Then $g^{\prime}(\mu_X)=1+\epsilon h^{\prime}(\mu_X),$ and $g(\mu_X)=1+\epsilon h(\mu_X)$. Plugging this into the first-order approximation of Lin’s CCC and expanding to second order in $\epsilon$ yields
\begin{equation}\label{eq:trans2}
\rho_c(X,g(X)) \approx 1 - \frac{\epsilon^2 h(\mu_X)^2}{2\sigma_X^2} - \frac{1}{2}\,\epsilon^2 \big(h'(\mu_X)\big)^2 + o(\epsilon^2).
\end{equation}
Thus, this illustrates how small deviations in both location and slope reduce concordance, offering a clear sensitivity analysis of Lin’s CCC with respect to minor nonlinear transformations.

In the univariate context, \cite{King:2001} emulated the construction of the CCC to derive a robust coefficient. 
Assuming that \( g(\cdot) \) is a distance measure, this robust coefficient is defined based on \( \mathbb{E}[g(X - Y)] \). 
If the vector \( (X, Y)^{\top} \) follows a bivariate distribution with cumulative distribution function \( F_{XY} \), then the coefficient is given by  
\begin{equation}\label{robust}
\rho_g = 
\frac{
[\mathbb{E}_{I}(g(X - Y)) - \mathbb{E}_{I}(g(X + Y))] 
- [\mathbb{E}_{F_{XY}}(g(X - Y)) - \mathbb{E}_{F_{XY}}(g(X + Y))]}
{\mathbb{E}_{I}(g(X - Y)) - \mathbb{E}_{I}(g(X + Y)) 
+ \tfrac{1}{2}\mathbb{E}_{F_X}(g(2X)) - \mathbb{E}_{F_Y}(g(2Y))},
\end{equation}
where \( \mathbb{E}_{I}(\cdot) \) denotes the expected value of \( g \) under independence. 
This generalization reduces to Lin’s CCC when \( g(z) = z^2 \).

When a sample of size \( n \) is available, say \( (X_i, Y_i), \, i = 1, \ldots, n, \) 
the sample counterpart of \( \rho_g \) is defined as  
\begin{equation}\label{sample_robust}
\widehat{\rho}_g = 
\frac{
\frac{1}{n}\sum_i \sum_j[g(X_i - Y_j) - g(X_i + Y_j)] 
- \sum_i[g(X_i - Y_i)) - g(X_i + Y_i)]}
{\frac{1}{n}\sum_i \sum_j [g(X_i - Y_j) -(g(X_i + Y_j)) 
+ \tfrac{1}{2}\sum_i[g(2X_i) -g(2Y_i)]}.
\end{equation}
The inference of \( \widehat{\rho}_g \) was approached from a U-statistics perspective, 
since \( \widehat{\rho}_g \) can be expressed as a ratio of U-statistics. 
The consistency and asymptotic distribution of this estimator were derived following \cite{Davis:1968}. Considering a general distance function of the form \( g(z) = |z|^{\delta} \) for \( |z| < z_0 \) requires a discussion on the most suitable values of \( \delta \) to be used in practice. 
In the study by \cite{King:2001}, Monte Carlo simulation experiments were conducted for this purpose, and the results indicated that the best-performing values of \( \delta \) were 1.0 and 1.5. Later,  \cite{Tashakor:2019} proposed an approach that extends the existing methods of robust estimators by focusing on functionals that yield robust $L$-statistics.

An alternative approach in this context was recently proposed by \cite{Vallejos:2025b}, 
in which the function \( g(z) = \lvert z \rvert \) was used to define a CCC-type estimator given by  
\begin{equation}\label{eq:L1}
  \rho_1 = 1 - \frac{\mathbb{E}(\lvert X - Y\rvert)}{\mathbb{E}(\lvert X - Y\rvert : \sigma_{XY} = 0)},
\end{equation}
where the distributional assumptions are the same as in \ref{Sigma}. 
The coefficient in \eqref{eq:L1} can be computed without the need to estimate nuisance parameters, 
as is often required in robustness analyses. 
Moreover, theoretical properties were established for both multivariate normal and elliptically contoured  distributions. 

If we assume that $(X, Y)^\top\sim \mathcal{N}(\bm{\mu},\bm{\Sigma})$, then 
\begin{equation}\label{eq:CCC1}
  \rho_1 = 1 -\frac{\gamma\big(1 - 2\Phi(-\gamma/\tau)\big) + \tau\sqrt{2/\pi}\exp\big(-\half(\gamma/\tau)^2\big)}
  {\gamma\big(1 - 2\Phi(-\gamma/\sqrt{\sigma_{X}^2 + \sigma_{Y}^2})\big) + \sqrt{2(\sigma_{X}^2 + \sigma_{Y}^2)/\pi}
  \exp\big(-\half\gamma^2/(\sigma_{X}^2 + \sigma_{Y}^2)\big)},
\end{equation}
where $\gamma = \mu_X - \mu_Y$, $\tau^2 = \sigma_{X}^2 + \sigma_{Y}^2 - 2\sigma_{XY}$ and $\Phi(\cdot)$ is the cumulative 
distribution function of the standard normal distribution. In particular, if $\mu_X = \mu_Y$, the coefficient \eqref{eq:CCC1} 
can be expressed as
\begin{equation}\label{eq:coef}
  \rho_1 = 1 - \sqrt{\frac{\sigma_{X}^2 + \sigma_{Y}^2 - 2\sigma_{XY}}{\sigma_{X}^2 + \sigma_{Y}^2}}.
\end{equation}

The $\rho_1$ coefficient can also be computed for elliptically contoured distributions. As an example, assume that  $\bm{X}\sim\mathsf{EC}_k(\bm{\mu},\bm{\Sigma};g)$. Then the probability density function is given by 
\[
  f(\bm{x};\bm{\mu},\bm{\Sigma}) = C_g \lvert\bm{\Sigma}\rvert^{-1/2} g[(\bm{x}-\bm{\mu})^\top
  \bm{\Sigma}^{-1}(\bm{x}-\bm{\mu})],
\]
where $g$ is known as the density generator function \citep{Fang:1990} and $C_g$ denotes a constant of integration. 

\begin{proposition}\label{prop:1}
  Assume that $(X,Y)^\top \sim \mathsf{EC}_2(\bm{\mu},\bm{\Sigma};g)$. Then 
  \begin{equation}\label{eq:CCC1-EC}
    \rho_1 = 1 - \frac{\displaystyle\gamma\Big(1 - 2C_g\int_{-\infty}^{-\alpha_1}g(r^2)\rd r\Big) - 2C_g\tau\int_{-\infty}^{-\alpha_1} 
    r g(r^2)\rd r}{\displaystyle\gamma\Big(1 - 2C_g\int_{-\infty}^{-\alpha_2}g(r^2)\rd r\Big) - 2C_g\sqrt{\sigma_{X}^2 
    + \sigma_{Y}^2}\int_{-\infty}^{-\alpha_2} r g(r^2)\rd r},
  \end{equation}
  where $\alpha_1 = \gamma/\tau$ and $\alpha_2 = \gamma/\sqrt{\sigma_{X}^2+\sigma_{Y}^2}$.
\end{proposition}

\cite{Vallejos:2025b} presents results that considerably simplify the expressions for the coefficient, 
particularly in the case where \( \mu_X = \mu_Y \). 
Moreover, these authors provide conditions under which the sample version of \( \rho_1 \) is asymptotically normal. 
Returning to the problem of selecting the optimal value of \( \delta \) that yields a robust, winsorized estimation, 
\cite{Vallejos:2025b} demonstrate that when an \( L_p \)--type distance function \( g(z) = |z|^p, \, p \in \mathbb{N}, \) is considered, 
the asymptotic variance of a plug-in version of \( \rho_p \), denoted \( \widehat{\rho}_p \), is minimized at \( p = 1 \). 
This result reinforces the idea that \( \rho_1 \) is not only robust but also possesses an optimality property 
that supports its practical use.

It is well established that traditional estimation methods for the concordance correlation coefficient (CCC) are not suitable for censored data and, consequently, cannot be directly applied to survival outcomes. To address this limitation, \cite{Guo:2007} proposed a nonparametric estimation approach for the CCC based on the bivariate survival function. The resulting estimator was shown to be strongly consistent and asymptotically normal, with a bootstrap-based variance estimator that is also consistent. Moreover, the authors introduced a time-dependent agreement coefficient, extending Lin’s  CCC framework to quantify the agreement between survival times among individuals who remain alive beyond a specified time point.

A general strategy for predicting a random variable \( Y \) from another variable \( X \) is to find a function of \( X \) that maximizes a chosen measure of association with \( Y \). Starting from Pearson’s correlation \cite{Bottai:2022} found a proposal that  leads to an infinite class of possible predictors. The least-squares predictor arises by equating their means and matching the variance of the predictor to that of the conditional expectation of \( Y \) given \( X \). If, instead, the predictor’s variance is set equal to that of \( Y \), the resulting  predictor maximizes Lin’s (1989) concordance correlation coefficient with \( Y \). \cite{Christensen:2022} showed that the optimal correlation-based predictor, under the mean and variance constraints examined by \cite{Bottai:2022}, also minimizes the expected squared prediction error when those same constraints are imposed on the predictors. Subsequently, \cite{Kim:2023} examined the distributional properties and predictive performance of the estimated maximum agreement linear predictor.

In a generalized linear mixed-effects models framework, \cite{Tsai:2018} suggested a variance components  approach that allows dependency between repeated measurements over time to assess intra-agreement for each observer and inter- and total agreement among multiple observers simultaneously under extended three-way generalized linear mixed-effects models for longitudinal
normal and Poisson data. The CCC index between two continuous variables \( X \) and \( Y \) for longitudinal repeated measurements, as introduced by \cite{King:2007}, is defined by incorporating a non-negative definite weighting matrix \( \bm{D} \) that accounts for the dependence between repeated measurements over time, such that
the  CCC in this case is defined as
\[
\rho_{c,rm} = 1 - 
\frac{
E\!\left[ ( \mathbf{X} - \mathbf{Y} )^{\mathsf{T}} 
\mathbf{D} ( \mathbf{X} - \mathbf{Y} ) \right]
}{
E_I\!\left[ ( \mathbf{X} - \mathbf{Y} )^{\mathsf{T}} 
\mathbf{D} ( \mathbf{X} - \mathbf{Y} ) \right]
},
\tag{3}
\]
where \( E_I[\cdot] \) denotes the expectation computed under the assumption of independence between observers. The idea of incorporating a matrix $\bm{D}$ into the CCC coefficient was also explored by \cite{Vallejos:2025} in the context of spatial statistics. This topic will be revisited in Section~\ref{sec:spatial}.

\cite{Pandit:2024} investigated the analytical relationship between the estimated CCC, the mean squared error (MSE), and the correlation coefficient. Given a random sample $(X_i, Y_i), \; i = 1, \ldots, n$, the MSE is defined as
\[
\text{MSE} = \frac{1}{n} \sum_{i=1}^n (X_i - Y_i)^2.
\]
They then derived the following relationship:
\[
\widehat{\rho}_c = \left( 1 + \frac{\text{MSE}}{2\widehat{\sigma}_{XY}} \right)^{-1},
\]
where $\widehat{\sigma}_{XY} = \frac{n-1}{n}S_{XY}$. The authors further demonstrated that, for a fixed MSE, there is no one-to-one correspondence with a single CCC value. Instead, a range of possible CCC values can occur, depending on the distribution of prediction errors. Moreover, they derived the maximum and minimum attainable CCC values for a given MSE, providing explicit formulas and graphical illustrations.

From a Bayesian perspective, \cite{Feng:2015} developed a  method for estimating Lin’s CCC   that remains accurate when the data are skewed, heavy-tailed, or asymmetric. Using skew-elliptical distributions, the approach models non-Gaussian measurement errors and can also incorporate covariates, missing data, and replicates. Simulations and real data confirm that the Bayesian skew-elliptical CCC outperforms normal-based CCC methods under asymmetric data.

\subsection{The probability of agreement}

\cite{Stevens:2017} introduced a probabilistic metric that expresses the likelihood that two measurements on the same subject fall within a predefined tolerance range. In contrast to correlation-based measures such as Lin’s CCC, which merge precision and accuracy into a single summary value, the Probability of Agreement (PA) explicitly quantifies how frequently two measurements coincide within a practically meaningful limit.

 Let $(X_{i}, Y_{i}), i=1,\ldots,n$ denote a random sample drawn from a bivariate normal distribution with mean vector $\bm{\mu}$ and covariance matrix $\bm{\Sigma}$. To assess the degree of agreement between $X$ and $Y$, we consider the pairwise differences
\[
D_i = X_{i} - Y_{i}, \hspace{3mm} i = 1, \ldots, n.
\]
The probability of agreement is defined as
\begin{equation}\label{eq:concordance}
\psi_c = P(|D_i| \leq c), \quad c > 0,
\end{equation}
where $c$ represents the maximum acceptable difference from a practical standpoint. The interval $(-c, c)$ is commonly referred to as the ``clinically acceptable difference'' (CAD).

Under the assumption of normality, the PA in Equation~\eqref{eq:concordance} can be expressed as
\begin{equation}\label{eq:concor2}
\psi_c = \Phi\left(\frac{c - \mu_D}{\sigma_D}\right) - \Phi\left(-\frac{c + \mu_D}{\sigma_D}\right),
\end{equation}
where $\Phi(\cdot)$ denotes the cumulative distribution function of the standard normal distribution, $\mu_D = \mu_X - \mu_Y$, and $\sigma_D^2 = \sigma_{X}^2 + \sigma_{Y}^2 - 2\sigma_{XY}$.
The threshold for considering the variables interchangeable depends on the practitioner; \cite{Stevens:2017} recommend using $\psi_c \geq 0.95$ as a general guideline.

Assuming normality, inference for $\bm{\mu}$ and $\bm{\Sigma}$ can be performed via maximum likelihood (ML) estimation \cite[][\S3.2]{Anderson:2003}. Substituting these ML estimates into Equation~\eqref{eq:concor2} yields the ML estimator of $\psi_c$, denoted by $\widehat{\psi}_c$. Under mild regularity conditions, \cite{Leal:2019} proved that $\widehat{\psi}_c$ is asymptotically normal, based on the asymptotic distribution of the ML estimators under normality and the delta method. Consequently, approximate confidence intervals and hypothesis tests for $\psi_c$ can be readily constructed.

The choice of $c$ is user defined and application-dependent, reflecting the practically acceptable difference between systems, and since the PA  is sensitive to this choice, different 
$c$ values can yield different results making it advisable to assess sensitivity or plot PA as a function of $c$.

\begin{figure}[htp]
    \centering
    \includegraphics[width=0.55\linewidth]{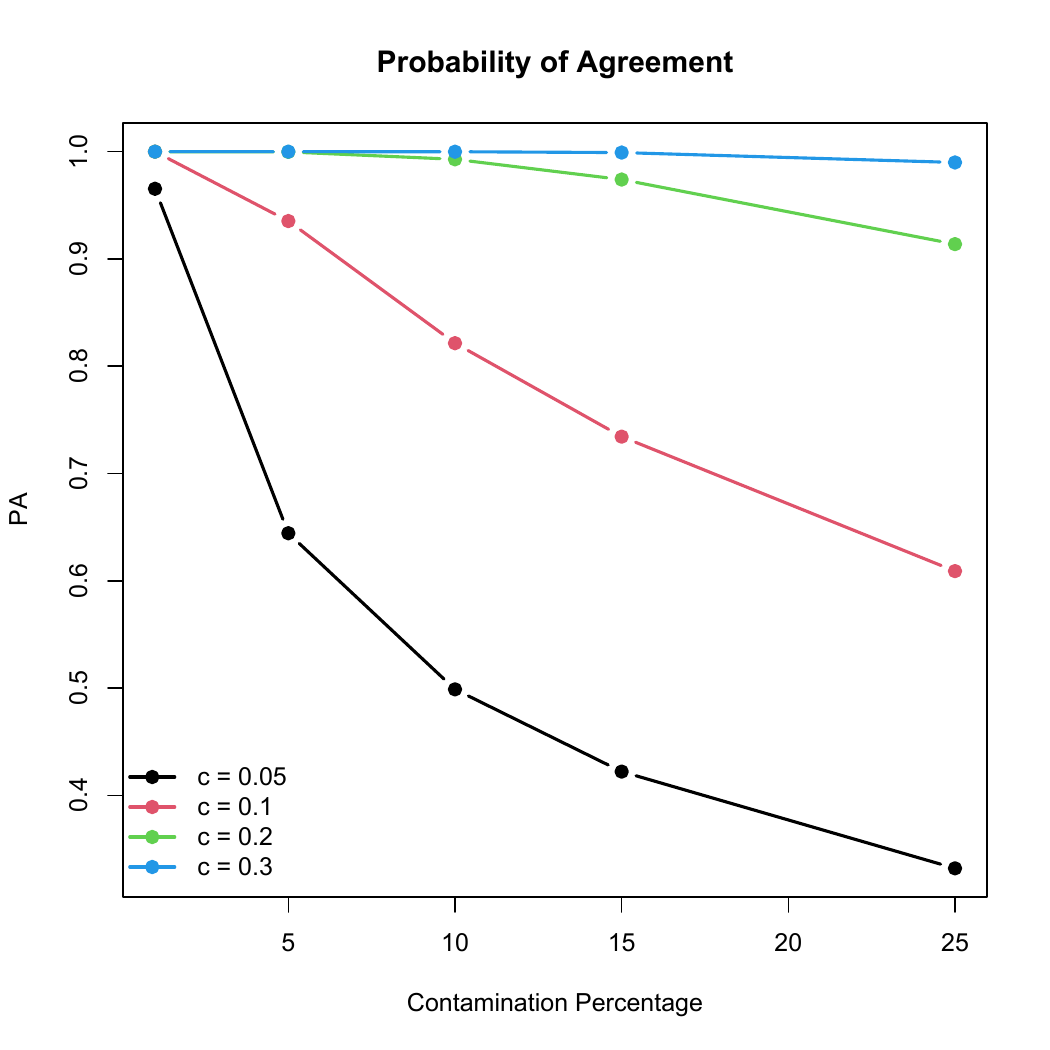}
    \caption{PA between image shown in Figure  \ref{fig:forest}(a) and images (c)-(f), for different values of $c.$}
    \label{fig:PA_plot}
\end{figure}

As an illustration, we compute the PA for the images displayed in Figure \ref{fig:forest}. For the values $c = \{0.05, 0.1, 0.2, 0.3\}$, the PA between the original image and the contaminated images with 1\%, 5\%, 10\%, 15\% and 25\% contamination was computed. Figure \ref{fig:PA_plot} shows how the PA curves decrease as the contamination percentage increases.

Although there are no explicit rules for selecting the value of $c$, we infer that this choice is related to the variance of the underlying process. A conjecture to be examined in future research is
\[
c \propto \sigma^{\alpha},
\]
where $\sigma$ is the standard deviation of the gold standard (when available) and $\alpha > 0$. For the forest data shown in Figure \ref{fig:forest}, this interpretation is reasonable because in our experiment $c \in [0.05,0.3]$, and the standard deviation of the original image is $0.1673987$. Hence, $\alpha = 1$ appears suitable for this example.

A number of applications have emerged following the seminal work of \cite{Stevens:2017}. For instance, \cite{Stevens:2018} extended the PA framework to situations where measurement error variances differ across the range of measurement, i.e., heteroscedastic measurement systems. 

\cite{Stevens:2020b} defined 
the PA over the entire domain of interest  as the probability that the two response values $Y_1(\mathbf{x})$ and $Y_2(\mathbf{x})$ are "close enough." The closeness is measured using a user-defined distance function $L$, typically a squared difference, and compared against a tolerance limit ($c$). The Bayesian PA is formally defined as the expected value of the agreement function, averaged over the posterior distributions of the model parameters ($\bm{\theta}_1, \bm{\theta}_2$) and the distribution of the design points ($\mathbf{x}$):$$ \text{PA} = E_{\mathbf{x}, \bm{\theta}_1, \bm{\theta}_2} \left[ I\left( L(Y_1(\mathbf{x}, \bm{\theta}_1), Y_2(\mathbf{x}, \bm{\theta}_2)) \leq c \right) \right]$$where $I(\cdot)$ is the indicator function.

\cite{deCastro:2021} extended the PA framework by shifting the focus from the comparison of modeled response surfaces, as in \cite{Stevens:2020b}, to the evaluation of measurement systems that generate repeated observations of an underlying true value. Whereas \cite{Stevens:2020b} quantify agreement between fitted surfaces across a design space, \cite{deCastro:2021} instead model the latent true quantity along with system-specific biases and variances, allowing them to assess how likely it is that two instruments will produce measurements falling within a specified practical tolerance when applied to the same true value. 

\cite{Taffe:2023} connects and extends the PA framework by showing how it aligns with clinical tolerance–limit and coverage–probability approaches commonly used to assess agreement in medical and biological measurement settings. The paper reframes agreement as the probability that two measurements on the same subject fall within clinically acceptable limits, integrating PA concepts with established methods such as Bland–Altman limits, coverage probabilities, and tolerance intervals. By doing so, Taffé provides a unified perspective in which agreement is interpreted through the probability of remaining within a prespecified clinical threshold, offering a bridge between statistical agreement metrics and practical decision-making in clinical measurement validation. \cite{Larsen:2024} suggested the PA as a statistically robust and practically intuitive alternative to the widely used Bland-Altman Limits of Agreement  for comparing two continuous measurement tools. While limits of agreement only describes the expected range of differences, the PA method introduces a model-based framework that directly addresses the question of interchangeability. By defining a user-specified tolerance ($c$) of practical equivalence, PA quantifies the probability that the difference between the two systems falls within this acceptable range, providing a single, actionable metric for quality engineers and clinical researchers to decide if the systems are functionally equivalent. \cite{Nadi:2024}  demonstrate how to explicitly incorporate the two primary components of measurement variation: repeatability (instrument error) and reproducibility (operator/appraiser error). By integrating these variance components directly into the PA calculation, the resulting agreement metric accurately reflects all sources of measurement system error, providing a more comprehensive and statistically sound assessment of whether two measurement systems can be considered functionally interchangeable for a given task. This ensures that the estimated PA is a reliable measure of interchangeability in real-world industrial settings.

\section{Measuring Agreement Over Time}
\label{sec:time}
Here, we present the approach proposed by \cite{Li:2005}, who introduced a measure for evaluating agreement between functional measurements—a type of data commonly encountered in medical research and various other scientific fields. Let $X(t)$ and $Y(t)$, $t \in T$ (a finite closed real interval), denote the measurements obtained from two instruments. Treating the random functions $X(\cdot)$ and $Y(\cdot)$ as elements of a functional probability space $\mathscr{T}$, \cite{Li:2005} defined the inner product
\begin{equation}\label{eq:inner}
\langle X(\cdot), Y(\cdot) \rangle = \mathbb{E}\!\int_T X(t)\,Y(t)\,w(t)\,dt,
\end{equation}
where $w(\cdot)$ is a non-negative weight function defined on $T$. As a consequence, a concordance correlation coefficient for the processes 
$X(\cdot)$ and $Y(\cdot)$ can be defined as
\begin{equation}\label{CCC_T}
\rho_c(X,Y) =
\frac{2\langle X - \mathbb{E}[X],\, Y - \mathbb{E}[Y]\rangle}
{\|\mathbb{E}[X] - \mathbb{E}[Y]\|^2 
 + \|X - \mathbb{E}[X]\|^2 
 + \|Y - \mathbb{E}[Y]\|^2},
\end{equation}
where $\mathbb{E}[X] = \mathbb{E}[X(t)]$ and 
$\mathbb{E}[Y] = \mathbb{E}[Y(t)]$ denote the mean functions of the respective processes.
This coefficient satisfies several desirable theoretical properties that link $\rho_c$ to the traditional correlation coefficient $\rho$, which can be defined using the same inner product introduced in \eqref{eq:inner}. Note that the weight function allows different regions of the domain $t$ 
to receive varying levels of importance. To estimate this weight function, one must first estimate the density of $t$ based on the observed values $t_1,\ldots,t_N$. The authors suggest using a kernel estimator of the form
\begin{equation}\label{eq:kernel}
\widehat{w}(t)=\frac{1}{Nh}\sum_{j=1}^N K\{(t_j-t)/h\},
\end{equation}
where $K(\cdot)$ is a kernel density function \citep[See for example,][p.131-138]{Wasserman:2006}.

For the inference of $\rho_c$, and without loss of generality, assume that $T = [0,1]$. Suppose that for subject $i$, $i = 1,\ldots,n$, the pair $(X_i(t), Y_i(t))$ is observed at time points $t = t_j$, $j = 1,\ldots,N$, with $0 \le t_1 < t_2 < \cdots < t_N \le 1$. Let $\Delta_j = t_{j+1} - t_j$ denote the gap between consecutive observation times. Then the sample counterpart of $\rho_c$ is given by
\begin{equation*}
\widehat{\rho}_c=\frac{2\sum_{i=1}^n \sum_{j=1}^N \{X_i(t_j)-\overline{X}(t_j)\}\{Y_i(t_j)-\overline{Y}(t_j)\}w(t_j)\Delta_j }{\sum_{j=1}^N \{\overline{X}(t_j)-\overline{Y}(t_j)\}^2 w(t_j)\Delta_j+\frac{1}{n}\sum_{i=1}^n \sum_{j=1}^N \left[ \{X_i(t_j)-\overline{X}(t_j)\}^2+\{Y_i(t_j)-\overline{Y}(t_j)\}^2  \right]w(t_j)\Delta_j},
\end{equation*}
where $\overline{X}(t_j)=\frac{1}{n}\sum_{i=1}^n X_i(t_j)$ and 
$\overline{Y}(t_j)=\frac{1}{n}\sum_{i=1}^n Y_i(t_j)$ are the sample means of $X(t_j)$ and $Y(t_j)$, respectively. \cite{Li:2005} established conditions under which the asymptotic normality of $\rho_c$ holds. Moreover, these authors extended the CCC to image data by embedding images into an appropriate functional space. A standard error formula for the resulting estimator was also derived and empirically evaluated.

An alternative approach to measure the agreement between two time series is the index describe by \cite{Vallejos:2008}. In this case the two time series have been measured on the same time points and there is no replicates for them. i.e., the only information available for both  processes is $(X(t),Y(t)), t=1,\ldots,n.$ This idea has some common ingredients with the CCC do functional time series but the construction of the coefficient is comparing  successive slopes on both series in time. In the literature this concept has been called comovement, which is no exactly an extension of the CCC, but it is a particular case of a well known coefficient used in spatial modelling, called codispersion coefficient \citep{Rukhin:2008}. For wide-sense stationarity differentiable processes, $X(t)$ and
$Y(t)$, such that $X^{\prime}(t)=X(t+1)-X(t),$ and
$Y^{\prime}(t)=Y(t+1)-Y(t),$ the comovement coefficient is defined
by the formula
$$\mbox{cm}(X(t),Y(t))=\frac{\mathbb{E}[X^{\prime}(t)Y^{\prime}(t)]}{[\mbox{Var}(X^{\prime}(t))\mbox{Var}(Y^{\prime}(t))]^{1/2}}.$$ 
It is assumed that $\mathbb{E}[X_t^{\prime}]< \infty$ and
$\mathbb{E}[Y_t^{\prime}]< \infty.$

For
two sequences $X(t)$ and $Y(t),$   the estimator of this
coefficient is the following

\begin{equation}\label{eq:comovement}
\widehat{\mbox{cm}}(X(t),Y(t))=\frac{\sum (X(t+1)-X(t))(Y(t+1)-Y(t))}
{\left(\sum (X(t+1)-X(t))^2 \sum (Y(t+1)-Y(t))^2  \right)^{1/2}
}.
\end{equation}

The comovement coefficient shares several standard properties with the correlation coefficient. It is easy to verify that the coefficient and its sampling variants is translation invariant, positively homogeneous, symmetric in its arguments, positive definite for a sequence and its lagged versions, and can be interpreted as the cosine of the angle between the vectors formed by the first differences of the sampled series. The only mildly nontrivial aspect is establishing positive definiteness, which can be shown by adapting the classical proof used for the autocovariance function \citep{McLeod:1984}.
It is also important to note that nothing in the definition of the comovement statistic requires the two sequences under study to be sampled at equal intervals, as is typically assumed in time series analysis. First differencing can be applied regardless of whether the observation spacings are uniform or continuous. The only essential requirement is that observations from the two sequences be temporally matched.

We emphasize that the comovement coefficient is not, strictly speaking, a measure of agreement between two sequences \(X(t)\) and \(Y(t)\). Rather, within the broader context of comparative analysis, comovement serves as a complementary measure that provides additional insight in the presence of serial dependence. Agreement coefficients alone cannot quantify the extent to which two series evolve or fluctuate in a coordinated manner. When used jointly, however, agreement and comovement measures offer a more comprehensive characterization of the temporal relationship between the processes.

The variance of the coefficient in \eqref{eq:comovement} is not straightforward to derive. \cite{Vallejos:2008} obtained the asymptotic variance for the case of two AR(1) processes; however, extending this result to higher-order models is not immediate. In that work, a block bootstrap procedure was also implemented to estimate the variability of the coefficient \eqref{eq:comovement}.

Additional notions of comovement between two time series can be found in \cite{Barberis:2005}, \cite{Baur:2003}, and \cite{Baba:2024}, among others. Comparisons based on phase-spectrum measures for time series can be found in \cite{Shumway:2006}.

Finally, we distinguish between agreement and Dynamic Time Warping (DTW) methods. DTW and statistical agreement measures address distinct methodological aims. Agreement metrics, such as Lin’s concordance correlation coefficient or the Bland–Altman framework assess pointwise equivalence between two instruments measuring the same construct on a common time grid, explicitly quantifying bias and scale differences. In contrast, DTW is an elastic alignment method that allows non-linear temporal deformations, matching observations from different time indices to minimize an overall alignment cost. As a result, a small DTW distance reflects similarity in shape under temporal warping, not equality of values on their original scale. Consequently, DTW cannot be interpreted as evidence of measurement agreement or interchangeability; rather, it characterizes pattern resemblance when timing differences are expected or tolerated \citep[see, e.g.,][Chapter 4]{Muller:2007}.

\section{Assessing the Agreement in  Multivariate Contexts}
\label{sec:mult}

In the previous sections, we have focused on agreement between two continuous variables. Nonetheless, more complex scenarios require appropriate agreement measures, particularly when responses are collected repeatedly by each rater or method.

For paired or unpaired repeated-measurement study designs, \cite{Chinchilli:1996} proposed a weighted CCC based on a random coefficient model that allows within-subject variances to vary across subjects. For each subject, the CCC is computed as an average of the CCCs associated with the least-squares random vectors. The overall CCC is then defined as a weighted average of the subject-specific coefficients, with weights determined by the magnitude of within-subject variability. Subsequently, \cite{King:2007} proposed another version of the CCC in the presence of repeated measurements. They characterized the amount of agreement between two $p \times1$ random vectors, say $\bm X$ and $\bm Y$ by using $\mathbb{E}[(\bm X-\bm Y)^{\top}\bm D (\bm X-\bm Y) ]$, where $\bm D$ is a $p\times p$ nonnegative definite matrix of weights among the different repeated measurements. As a result, the CCC was defined through
\begin{align}\label{coef:king}
    \rho_c  &= 1 - \frac{\mathbb{E}[(\bm{X} - \bm{Y})^{\top} \bm{D}(\bm{X} - \bm{Y})]}{\mathbb{E}[(\bm{X} - \bm{Y})^{\top} \bm{D}(\bm{X} - \bm{Y})\mid\bm{\Sigma}_{XY}=\bm{0}]} \notag\\
    &= \frac{\operatorname{Tr}[\bm{D} \bm{\Sigma}_{XY} + \bm{D} \bm{\Sigma}_{XY}^{\top}]}{\operatorname{Tr}[\bm{D} \bm{\Sigma}_{X} + \bm{D} \bm{\Sigma}_{Y}] + \left(\bm{\mu}_X - \bm{\mu}_Y\right)^{\top} \bm{D}\left(\bm{\mu}_X - \bm{\mu}_Y\right)},
\end{align}
where the $2p \times 1$ random vector $(\boldsymbol{X}^\top, \boldsymbol{Y}^\top)^\top$ is assumed to follow a Gaussian 
distribution with mean vector 
$(\boldsymbol{\mu}_X^\top, \boldsymbol{\mu}_Y^\top)^\top$ and covariance matrix
\[
\begin{pmatrix}
\boldsymbol{\Sigma}_{X} & \boldsymbol{\Sigma}_{XY} \\
\boldsymbol{\Sigma}_{YX} & \boldsymbol{\Sigma}_{Y}
\end{pmatrix}.
\]

\cite{Carrasco:2009} proposed a CCC for longitudinal repeated measurements by appropriately defining the intraclass correlation coefficient within a variance-components linear mixed-effects model. They demonstrated that this CCC coincides with the repeated-measures CCC introduced by \cite{King:2007} when  $\bm D$ is the identity matrix. Subsequently, \cite{Hiriote:2011} proposed a new repeated-measures CCC that can be shown to satisfy the properties required to quantify overall agreement between two 
$p\times 1$ vectors of random variables, while also offering greater intuitive appeal than earlier approaches. 

To quantify the overall agreement between two vectors, the matrix $\boldsymbol{M}_{\rho_c}$ is defined as
\[
\boldsymbol{M}_{\rho_c}
= \boldsymbol{I}_p
- \boldsymbol{V}_{I}^{-1/2}\,\boldsymbol{V}_D\,\boldsymbol{V}_I^{-1/2},
\]
where $\boldsymbol{V}_D = \mathbb{E}\!\left[(\boldsymbol{X}-\boldsymbol{Y})(\boldsymbol{X}-\boldsymbol{Y})^{\top}\right]$
and
$\boldsymbol{V}_I = \mathbb{E}\!\left[(\boldsymbol{X}-\boldsymbol{Y})(\boldsymbol{X}-\boldsymbol{Y})^{\top} \mid \text{independence}\right]$
denote the covariance matrices of $\boldsymbol{X}-\boldsymbol{Y}$ under dependence and independence, respectively, and $\boldsymbol{I}_p$ is the $p \times p$ identity matrix.We note that the matrix $\boldsymbol{M}_{\rho_c}$ constitutes a generalization of Lin's CCC, since for $p = 1$ it reduces exactly to Lin's CCC. As a consequence, a matrix-based CCC is defined through
\begin{equation}\label{mult_CCC}
\rho_g=1-\frac{g(\bm I-\boldsymbol{M}_{\rho_c})}{g(\bm I)}=1-\frac{g(\boldsymbol{V}_{I}^{-1/2}\,\boldsymbol{V}_D\,\boldsymbol{V}_I^{-1/2})}{g(\bm I)},
\end{equation}
where $g(\cdot)$ is a matrix norm. For instance, the Frobenius norm is a natural and appealing choice in our setting. For a matrix $\boldsymbol{A}$, it is defined as
\[
g(\boldsymbol{A}) = \operatorname{tr}\!\left(\boldsymbol{A}^{\top}\boldsymbol{A}\right).
\]
If  $(\bm X_i, \bm Y_i)$, $i=1,2,\ldots,n$ are i.i.d. random vectors from 
$2p$-variate distribution with finite fourth moments, then a sample version of $\rho_g$ is
\begin{equation}\label{estim:rho_g}
\widehat{\rho}_g=1-\frac{g(\widehat{\boldsymbol{V}}_{I}^{-1/2}\,\widehat{\boldsymbol{V}}_D\, \widehat{\boldsymbol{V}}_I^{-1/2})}{g(\bm I)},
\end{equation}
where 
$
\widehat{V}_D
=
\frac{1}{n}
\sum_{i=1}^n
(X_i - Y_i)(X_i - Y_i)^{\top},
$
and
$
\widehat{V}_I
=
\frac{1}{n(n-1)}
\sum_{i \neq j}
(X_i - Y_j)(X_i - Y_j)^{\top}.
$
\cite{Hiriote:2011} established the asymptotic distribution of $\widehat{\rho}_g$ and conducted Monte Carlo simulation studies to examine its finite-sample behavior. In addition, real data were analyzed to provide further insight into the practical performance of the proposed methodology. The two datasets considered—blood draw data and body fat data—illustrate cases of high overall agreement and moderate agreement between two sets of measurements, respectively.

\subsection{Critical Perspectives}

\cite{Leal:2019} introduced a local influence diagnostic framework for agreement analysis, extending Cook’s \citep{Cook:1986} methodology to assess the sensitivity of agreement measures to small perturbations in the data. Focusing on Lin’s concordance correlation coefficient and the probability of agreement, the authors formulated these indices as smooth functions of model parameters estimated under a parametric framework, allowing influence to be quantified through curvature-based diagnostics. Their results show that observations exerting substantial influence on agreement measures are not necessarily influential for parameter estimation or model fit, highlighting the importance of dedicated influence diagnostics in method comparison studies. Simulation studies and real-data applications illustrate the practical utility of the approach in identifying observations that disproportionately affect agreement conclusions.

Despite the advancement of such influence-detection methods, the CCC remains a point of contention; for instance, \cite{Atkinson:1997} raised significant questions regarding the index's fundamental reliability in the face of varying sample characteristics. These authors delineated their concerns around two main points. First, the CCC is highly sensitive to between-subject variability: a heterogeneous sample may yield a high CCC even in the presence of poor agreement, whereas a homogeneous sample may result in a low CCC despite comparable levels of agreement. This dependence can lead to potentially misleading conclusions when assessing agreement. Second, a single CCC value does not disentangle the sources of disagreement, such as systematic bias, random error, or a restricted measurement range. As a consequence, the CCC provides limited diagnostic insight into how measurement quality might be improved, particularly when compared with approaches that explicitly separate these components. Based on these arguments, the authors discouraged the use of correlation-based measures—including Pearson’s correlation, intraclass correlation, and the CCC—for assessing agreement between methods or repeated measurements, and instead advocated the use of the Bland–Altman approach.

In a rejoinder, Lin and Chinchilli argued that the dependence of the CCC on the measurement range should not be regarded as a limitation, but rather as an inherent property shared by all correlation-based measures. Agreement evaluated over a narrow range cannot be meaningfully extrapolated to a broader domain; thus, the CCC appropriately reflects agreement within the intended analytical range. Moreover, they emphasized that the CCC is designed to provide a global summary of agreement over a specified range and should be interpreted in conjunction with graphical tools and limits of agreement. In this sense, no single index is universally sufficient, but the CCC remains a valid and informative measure when comparisons are conducted over comparable ranges.

\cite{Wadoux:2024}  criticized Lin's CCC as a standalone validation statistic in environmental modeling due to three key limitations. It does not differentiate the influence of bias versus correlation, its values are not comparable across various datasets or studies, and it shares the drawbacks of other linear correlation statistics. For comprehensive validation of models and maps, calculating multiple statistics alongside the CCC is recommended to represent different quality aspects, which can be visualized using diagrams like Taylor or solar plots. 
\section{Developments for Spatial Data}
\label{sec:spatial}

\subsection{The Spatial CCC Coefficient}

Recently, several methodological developments have been proposed to assess agreement in spatial data. It is important to note that no single agreement measure can be universally defined for all types of spatial data, as the appropriate choice depends on the nature of the observations, the context of the problem under investigation, and the data acquisition process. 

A first extension of Lin's CCC was introduced by \cite{Vallejos:2020}, who proposed a spatial formulation to quantify agreement between spatially indexed data, with particular application to image analysis. Their approach incorporates spatial dependence by defining a concordance measure based on pairs of observations separated by a given spatial lag  $\bm h$, allowing agreement to be evaluated as a function of spatial scale. The resulting spatial concordance correlation coefficient captures both marginal agreement and spatial autocorrelation, thereby distinguishing true agreement from similarity induced by spatial structure alone. The authors studied the theoretical properties of the proposed index, including its relationship to the classical CCC, and developed estimation procedures under both parametric and nonparametric frameworks. Simulation studies and real-data applications demonstrate that accounting for spatial dependence can substantially alter agreement assessment compared with non-spatial measures. Precisely, consider 
 $\bm Z(\bm s)=(X(\bm s), Y(\bm s))^{\top}$ being a bivariate second-order stationary random field  with  $\bm s, \bm h \in \mathbb{R}^2$, mean $(\mu_X,\mu_Y)^{\top}$, and covariance function
$$C(\bm h)=\left(\begin{matrix} C_X(\bm h) & C_{XY}(\bm h)\\
C_{YX}(\bm h) & C_{Y}(\bm h)
\end{matrix} \right),$$
where 
 $C_{X}(\bm h) = \text{Cov}[X(\bm s),X(\bm s+\bm h)],$ 
 $C_{Y}(\bm h) = \text{Cov}[Y(\bm s),Y(\bm s+\bm h)],$  and
 $C_{XY}(\bm h)= C_{YX}(\bm h)= \text{Cov}[X(\bm s),Y(\bm s+\bm h)].$
Then the spatial concordance correlation coefficient (SCCC) is defined as 
\begin{align} \label{eq:spatial_cor}
\rho_c(\bm h)&=1-\frac{\mathbb{E}[(X(\bm s+\bm h)-Y(\bm s))^2]}{\mathbb{E}[(X(\bm s+\bm h)-Y(\bm s))^2|C_{XY}(\bm 0)=0]} \notag\\
&=\frac{2C_{XY}(\bm h)}{C_{X}(\bm 0)+C_{Y}(\bm 0)+(\mu_X-\mu_Y)^2}.
\end{align}
This definition was supported by an analysis of its theoretical properties, complemented by simulation studies and a real-data application in which two cameras were compared using images of the same scene (forest image).  The dependence of this coefficient on $\boldsymbol{h}$ can be viewed as analogous to the semivariogram in spatial statistics, which further highlights the need to investigate the behavior of $\rho_c(\boldsymbol{h})$ within the class of isotropic processes. However, if one considers a coefficient that does not depend on $\boldsymbol{h}$ while still incorporating spatial autocorrelation, it becomes more convenient, as it avoids the need for a prior selection of a suitable value of $\boldsymbol{h}$. This can be achieved, for instance, by considering a kernel-smoothed version of $\rho_c(\boldsymbol{h})$, which incorporates spatial interactions through appropriate weighting schemes \citep[see, e.g.,][]{Fotheringham2002}. In this case, the only additional parameter is the bandwidth, which can be selected using standard techniques from nonparametric inference \citep{Wand:1995}.

Another important aspect of the SCCC concerns the limiting distribution of the sample counterpart of $\rho_c(\boldsymbol{h})$. Let $\bm Z(\bm s) = (X(\bm s), Y(\bm s))^\top$, $\bm s \in D \subset \mathbb{R}^2$, denote a Gaussian process with mean $\bm \mu = (\mu_X, \mu_Y)^\top$ and covariance function $\bm C(\bm h)$, $\bm h \in \mathbb{R}^2$. A plug-in estimator of the SCCC is given by
\begin{equation}\label{eq:concor}
\widehat{\rho}_c(\bm h) = \widehat{\eta}\, \widehat{\rho}_{XY}(\bm h),
\end{equation}
where
\[
\widehat{\eta} = \left( \frac{\widehat{v} + \widehat{v}^{-1} + \widehat{u}^2}{2} \right)^{-1}, 
\quad 
\widehat{v} = \left( \frac{\widehat{C}_{X}(\bm 0)}{\widehat{C}_{Y}(\bm 0)} \right)^{1/2}, 
\quad 
\widehat{u} = \frac{\widehat{\mu}_X - \widehat{\mu}_Y}{\left(\widehat{C}_{X}(\bm 0)\widehat{C}_{Y}(\bm 0)\right)^{1/4}}.
\]
Here, $\widehat{\mu}_X$, $\widehat{\mu}_Y$, $\widehat{C}_{X}(\bm 0)$, and $\widehat{C}_{Y}(\bm 0)$ denote the maximum likelihood estimators of $\mu_X$, $\mu_Y$, $C_{X}(\bm 0)$, and $C_{Y}(\bm 0)$, respectively.

The asymptotic normality of estimators of the form \eqref{eq:concor} has been extensively studied under increasing-domain asymptotics \citep{Mardia:1984}. Specifically, consider a process $\bm Z(\bm s)$ observed on a lattice $D_n \subset \Delta \mathbb{Z}^2$, with $0 < \Delta < \infty$, where the domains satisfy $D_n \subset D_{n+1}$. Within this framework, \cite{Vallejos:2020} established the asymptotic normality of $\widehat{\rho}_c(\bm h)$ using the delta method for the Wendland--Gneiting class of bivariate covariance functions \citep{Porcu:2015}. 

Although these results extend key properties of Lin's CCC to the spatial setting, they present limitations in image analysis applications. In particular, it is unrealistic to assume that a single global spatial covariance model adequately captures the heterogeneous textures typically observed in large images. To address this issue, \cite{Vallejos:2020} proposed a local strategy that partitions the image into non-overlapping windows and fits separate spatial models within each subregion. 

Beyond parametric SCCC formulations for geostatistical data, nonparametric approaches offer a flexible alternative for large-scale images, avoiding the need to estimate a large number of covariance parameters. A natural strategy for assessing concordance in images is to employ kernel-based estimators that combine a suitable discrepancy measure between pixel intensities with kernel weights to account for local spatial interactions \citep[e.g.,][]{Datta:2016}.

\subsection{The PA for Spatial Processes}
In this section, we describe the PA for georeferenced variables first introduced by \cite{Acosta:2024}. Let
$
\bm Z(\bm s) = \bigl(X(\bm s), Y(\bm s)\bigr)^{\top},  \bm s \in \mathcal{D} \subset \mathbb{R}^2,
$
be a bivariate second-order stationary random field with mean $(\mu_X, \mu_Y)^{\top}$ and covariance function
\[
\bm C(\bm h) =
\begin{pmatrix}
C_X(\bm h) & C_{XY}(\bm h) \\
C_{YX}(\bm h) & C_Y(\bm h)
\end{pmatrix}, \qquad \bm s, \bm h \in \mathbb{R}^2,
\]
where $C_X(\bm h) = \text{cov}\{X(\bm s), X(\bm s + \bm h)\}$, $C_Y(\bm h) = \text{cov}\{Y(\bm s), Y(\bm s + \bm h)\}$, and $C_{XY}(\bm h) = C_{YX}(\bm h) = \text{cov}\{X(\bm s), Y(\bm s + \bm h)\}$.

We define the spatially shifted difference
\begin{equation}\label{eq:spatdiff}
D(\bm s, \bm h) = X(\bm s) - Y(\bm s + \bm h),
\end{equation}
which quantifies the discrepancy between the two processes separated by a spatial lag $\bm h$. When $\bm h = \bm 0$, this reduces to the usual pointwise difference $X(\bm s) - Y(\bm s)$; however, in this case, the resulting PA fails to capture spatial dependence explicitly.

Assume that $\bm Z(\bm s)$ follows a Gaussian process with mean $\bm \mu = (\mu_X, \mu_Y)^{\top}$ and covariance function $\bm C(\bm h)$, $\bm h \in \mathcal{D}$. Then
\[
D(\bm s, \bm h) \sim \mathcal{N}\bigl(\mu_D, \sigma_D^2(\bm h)\bigr),
\]
where $\mu_D = \mu_X - \mu_Y$ and
$\sigma_D^2(\bm h) = C_X(\bm 0) + C_Y(\bm 0) - 2 C_{XY}(\bm h).$
Accordingly, the PA between $X(\bm s)$ and $Y(\bm s + \bm h)$ is defined as
\begin{equation}\label{eq:spatconcordance}
\psi_c(\bm h) = P \bigl( |D(\bm s, \bm h)| \leq c \bigr),
\end{equation}
which admits the closed-form expression
\begin{equation}\label{eq:psi_c}
\psi_c(\bm h)
= \Phi\!\left( \frac{c - \mu_D}{\sigma_D(\bm h)} \right)
 - \Phi\!\left( -\frac{c + \mu_D}{\sigma_D(\bm h)} \right),
\end{equation}
where $\Phi(\cdot)$ denotes the standard normal distribution function.

When $X(\bm s)$ and $Y(\bm s)$ are isotropic processes, so that $\sigma_D(\bm h)$ depends only on $\|\bm h\|$, the probabilities in \eqref{eq:spatconcordance} and \eqref{eq:psi_c} can be written as $\psi_c(\bm h) = \psi_c(\|\bm h\|)$. In this setting, the PA can be represented as a function of the spatial lag norm, in direct analogy with the covariance function of isotropic spatial processes. The PA defined in \eqref{eq:concordance} is then recovered as a special case of \eqref{eq:spatconcordance}, since $\psi_c = \psi_c(\bm 0)$.

Several properties that are immediate for other agreement measures require explicit derivation in this framework. For example, \cite{Acosta:2024} established the monotonicity of $\psi_c(\|\bm h\|)$ under isotropy. In particular, they showed that the spatial PA is a decreasing function of $\|\bm h\|$ for both the bivariate Matérn and the Gneiting--Wendland covariance models. Inference is conducted under the assumption that consistent estimators $\widehat{\mu}_D$ and $\widehat{\boldsymbol{\theta}}$ are available. Within an increasing-domain asymptotic framework, the limiting distribution of $\psi_c(\bm h)$ has been derived, which directly enables the construction of approximate hypothesis tests of the form
\[
\text{H}_0: \psi_c(\bm h;\mu_D,\boldsymbol{\theta}) = \psi_c^{(0)},
\qquad 0 \le \psi_c^{(0)} \le 1,
\]
against one of the following alternatives:
$\text{H}_1: \psi_c(\bm h;\mu_D,\boldsymbol{\theta}) \neq (> \text{or} <) \ \psi_c(\bm 0).$ This testing procedure is of practical relevance, as it allows the PA at a fixed spatial lag $\bm h$ to be formally compared with the nominal benchmark value proposed by \cite{Stevens:2017}.

Extensions of the spatial PA to spatiotemporal settings have also been investigated for Gaussian processes under both separable and non-separable covariance structures. Numerical experiments were conducted to examine the sensitivity of the spatial PA to covariance separability and to provide guidance on the choice of the threshold parameter $c$, which is closely tied to the scale of the data. An application to forest images observed over time illustrated how past spatial realizations influence the temporal trend, offering insights that may be valuable for trend modeling and the development of more refined spatiotemporal models.

The computational efficiency of the composite likelihood approach used in \cite{Acosta:2024} warrants careful consideration. The implementations used in that study performed well for images of approximately 
$50\times 50$ pixels; however, the original images had to be rasterized and downscaled by two orders of magnitude before the relevant parameters could be estimated. Overcoming memory limitations to enable parameter estimation for substantially larger images (exceeding $10^6$ pixels) will require the development of innovative and highly parallelizable algorithms.

\subsection{A Spatial Coefficient for lattice Data}
Here we describe the proposal introduced by \cite{Vallejos:2025}, which presents a novel coefficient for quantifying agreement between two lattice sequences observed over the same areal units. The methodology is motivated by the comparison of alternative approaches for measuring poverty rates in Chile.

To construct a spatial concordance coefficient for lattice data, we rely on the generalized concordance coefficient proposed by \cite{King:2007}, under the assumption that the lattice observations are suitably modeled through a $\mathrm{GMCAR}$ framework \citep{Jin:2005}.

    Let $\bm{X}=(\bm{X}_1^{\top},\bm{X}_2^{\top})^{\top}$ denote a $2n\times 1$ random vector following a $2n$-variate Gaussian distribution. Suppose that $\bm X$ is generated from a $\mathrm{GMCAR}(\rho_1,\rho_2,\eta_0,\eta_1,\tau_1,\tau_2)$ model, where $\bm{X}_i=(X_{i1},\ldots, X_{in})^{\top}$ represents an $n\times 1$ random vector. Specifically,
    $$
    \begin{pmatrix}
        \bm{X}_1 \\
        \bm{X}_2 
    \end{pmatrix} \sim \mathcal{N}\hspace{-2pt}\left(\begin{pmatrix}
        \bm{\mu}_{1} \\
        \bm{\mu}_{2}
        \end{pmatrix},\begin{pmatrix}
    \bm{\Sigma}_{11} & \bm{\Sigma}_{12} \\
    \bm{\Sigma}_{12}^{\top} & \bm{\Sigma}_{22}
    \end{pmatrix}\right),
    $$
    where $\mathbb{E}[\bm X_i]=\bm{\mu}_{i}$ for $i=1,2$, and
    \begin{align*}
    &\bm{\Sigma}_{11}=\left[\tau_1\left(\bm{D}_w-\rho_1 \bm{W}_1\right)\right]^{-1}+\left(\eta_0 \bm{I}_{n}+\eta_1 \bm{W}_1\right)\left[\tau_2\left(\bm{D}_w-\rho_2 \bm{W}_1\right)\right]^{-1}\left(\eta_0 \bm{I}_{n}+\eta_1 \bm{W}_1\right),\\
    &\bm{\Sigma}_{12}=\left(\eta_0 \bm{I}_{n}+\eta_1 \bm{W}_1\right)\left[\tau_2\left(\bm{D}_w-\rho_2 \bm{W}_1\right)\right]^{-1},\\
    &\bm{\Sigma}_{22}=\left[\tau_2\left(\bm{D}_w-\rho_2 \bm{W}_1\right)\right]^{-1}.
    \end{align*}
    The spatial concordance coefficient, $\rho_{s,c}$, is defined as
    \begin{equation}\label{lattice_coef}
    \rho_{s,c}=\frac{\operatorname{Tr}[\bm{J}_n \bm{\Sigma}_{12}+\bm{J}_n \bm{\Sigma}_{12}^{\top}]}{\operatorname{Tr}[\bm{J}_n \bm{\Sigma}_{11}+\bm{J}_n \bm{\Sigma}_{22}]+\left(\bm{\mu}_1-\bm{\mu}_2\right)^{\top} \bm{J}_n\left(\bm{\mu}_1-\bm{\mu}_2\right)},
    \end{equation}
    where $\bm{J}_n=\bm{1}_n\bm{1}_n^{\top}$.

Because the estimated covariance matrix of the process $\bm X$ incorporates both inter- and intra-observation dependence through spatial autocorrelation and linking parameters, the matrix $\bm D$ in \eqref{lattice_coef} is taken to be $\bm J_n$.

Building on the framework of \cite{Jin:2005}, our proposal adopts a Bayesian inferential approach. We note that the asymptotic theory developed for CCC-type coefficients is not directly applicable to lattice models, as neither a clear increasing-domain scheme nor a well-defined infill asymptotic structure is available for lattice-based observations.

In order to give the elements used in the inference of the GMCAR process,  assume that $\bm X$ follows a $\mathrm{GMCAR}(\rho_1,\rho_2,\eta_0,\eta_1,\tau_1,\tau_2)$ process. Then,
\begin{equation*}
    \bm{X}_1 \mid \bm{X}_2 \sim  \mathcal{N}\left(\bm{\mu}_1+\left(\eta_0 \bm{I}_{n}+\eta_1 \bm{W}_1\right)\left(\bm{X}_2- \bm{\mu}_2\right),\left[\tau_1\left(\bm{D}_w-\rho_1 \bm{W}_1\right)\right]^{-1}\right),
\end{equation*}
and $\bm{X}_2 \sim \mathcal{N}\left(\bm{\mu}_2,\left[\tau_2(\bm{D}_w-\rho_2 \bm{W}_1\right)\right]^{-1})$.
Therefore, the joint distribution of $\bm{X}$ is given by
\begin{align*}
 f(\bm{X}\mid \bm{\mu}, \bm{\tau}, \bm{\rho}, \bm{\eta}) &\propto \tau_1^{n / 2}\operatorname{det}(\bm{D}_w-\rho_1 \bm{W}_1)^{1 / 2} 
  \exp \left\{-\frac{\tau_1}{2}\left[\bm{X}_1-\bm{\mu}_1-\left(\eta_0 \bm{I}_{n}+\eta_1 \bm{W}_1\right)\left(\bm{X}_2-\bm{\mu}_2\right)\right]^{\top}\right. \\
& \times\left(\bm{D}_w-\rho_1 \bm{W}_1\right)\left[\bm{X}_1-\bm{\mu}_1-\left(\eta_0 \bm{I}_{n}+\eta_1 \bm{W}_1\right)\left(\bm{X}_2- \bm{\mu}_2\right)\right]\Big\} \\
& \times \tau_2^{n / 2}\operatorname{det}(\bm{D}_w-\rho_2 \bm{W}_1)^{1 / 2} 
 \times \exp \left\{-\frac{\tau_2}{2}\left(\bm{X}_2-\bm{\mu}_2 \right)^{\top}\left(\bm{D}_w-\rho_2 \bm{W}_1\right)\left(\bm{X}_2-\bm{\mu}_2 \right)\right\},
\end{align*}
where $\bm{\mu}=(\bm{\mu}_1^{\top},\bm{\mu}_2^{\top})^{\top}$, $\bm{\tau}=(\tau_1,\tau_2)^{\top}$, $\bm{\eta}=(\eta_0,\eta_1)^{\top}$, and $\bm{\rho}=(\rho_1,\rho_2)^{\top}$. Consequently, the posterior distribution is given by
\begin{equation*}
    f(\bm{\mu}, \bm{\tau}, \bm{\rho}, \bm{\eta}\mid \bm{X})\propto f(\bm{X}\mid \bm{\mu}, \bm{\tau}, \bm{\rho}, \bm{\eta}) f(\bm{\mu}) f(\bm{\tau}) f(\bm{\rho}) f(\bm{\eta}),
\end{equation*}
provided that the priors are independent. The distribution of the spatial concordance coefficient defined in \eqref{lattice_coef} is subsequently derived via a plug-in estimator
\begin{equation}\label{bay_est}
\widehat{\rho}_{s,c}=\rho_{s,c}(\widehat{\bm{\theta}}),
\end{equation}
where $\bm{\theta}=(\bm{\mu},\bm{\tau},\bm{\eta},\bm{\rho})^{\top}$.

The authors in \cite{Vallejos:2025} applied the proposed spatial concordance coefficient to assess agreement between two methodologies for estimating poverty rates in Chile: the Horvitz–Thompson (HT) estimator and Small Area Estimation (SAE). Using county-level data from the 2011 CASEN survey \citep{Casas:2016}, the analysis focuses on lattice data from the Santiago Metropolitan and Valparaíso regions.
A bivariate GMCAR model is fitted within a Bayesian framework to capture both spatial autocorrelation within each method and cross-dependence between methods. Competing specifications incorporating different neighborhood orders are compared via DIC, and posterior samples are used to construct the plug-in distribution of the spatial concordance coefficient with HPD intervals.
Results indicate moderate agreement between HT and SAE in the Santiago Metropolitan Region, while substantially weaker and more uncertain agreement is observed in Valparaíso. These findings align with visual map comparisons and suggest that local spatial structure and small-area sampling variability play a critical role in method agreement. In addition,  the application demonstrates how the proposed coefficient provides a principled, model-based measure of agreement for areal data, with direct relevance for evaluating policy-relevant socioeconomic indicators.

\section{An Application}
\label{sec:application}
We present an application in the context of air quality monitoring, building on a dataset originally collected and documented by \cite{Thu:2023}. We do not further analyze the forest data introduced in Section \ref{motivating}, as a comprehensive study of that dataset is available in \cite{Vallejos;2018}.

The air quality monitoring dataset is relevant as it originates from a TIGA-funded initiative by the Rouen Normandy Metropolis aimed at improving urban mobility while reducing the environmental impact of transportation infrastructure. Within this framework, monitoring NO$_2$ emissions along major roadways—predominantly attributable to traffic—was identified as a priority, as reaffirmed by the European Environment Agency in 2022.

Low-cost sensors, which are substantially more affordable than reference monitoring devices—by up to two orders of magnitude—are planned for deployment in Rouen and its surrounding areas by ATMO Normandie, the regional Air Quality Monitoring Agency. Prior to their use in real-world conditions, these sensors require calibration against more reliable reference stations. The dataset reported by \cite{Thu:2023} comprises measurements from a subset of nine low-cost sensors colocated with reference monitoring devices over a five-month period characterized by elevated NO$_2$ concentrations. The primary objective of this dataset is to enable the calibration of low-cost sensors using established monitoring stations as the reference standard.

Measurements were originally acquired at one-minute and fifteen-minute intervals for low-cost sensors and monitoring stations, respectively, and transmitted via the cellular network; however, the dataset consists of hourly-averaged observations, which are the most commonly used. Data span the period from October 20, 2021, to March 25, 2022, yielding 29,605 observations. This interval corresponds to the period of highest annual NO$_2$ concentrations in Rouen. The raw datasets described above are publicly available at the following website: 
\url{https://data.mendeley.com/datasets/82dnstrd93/1}.  Previous work on the assessment of air quality microsensors relative to reference methods can be found in \cite{Borrego:2016}.

\begin{figure}[htbp]
    \centering
    \begin{subfigure}[b]{0.48\textwidth}
        \centering
        \includegraphics[width=\textwidth]{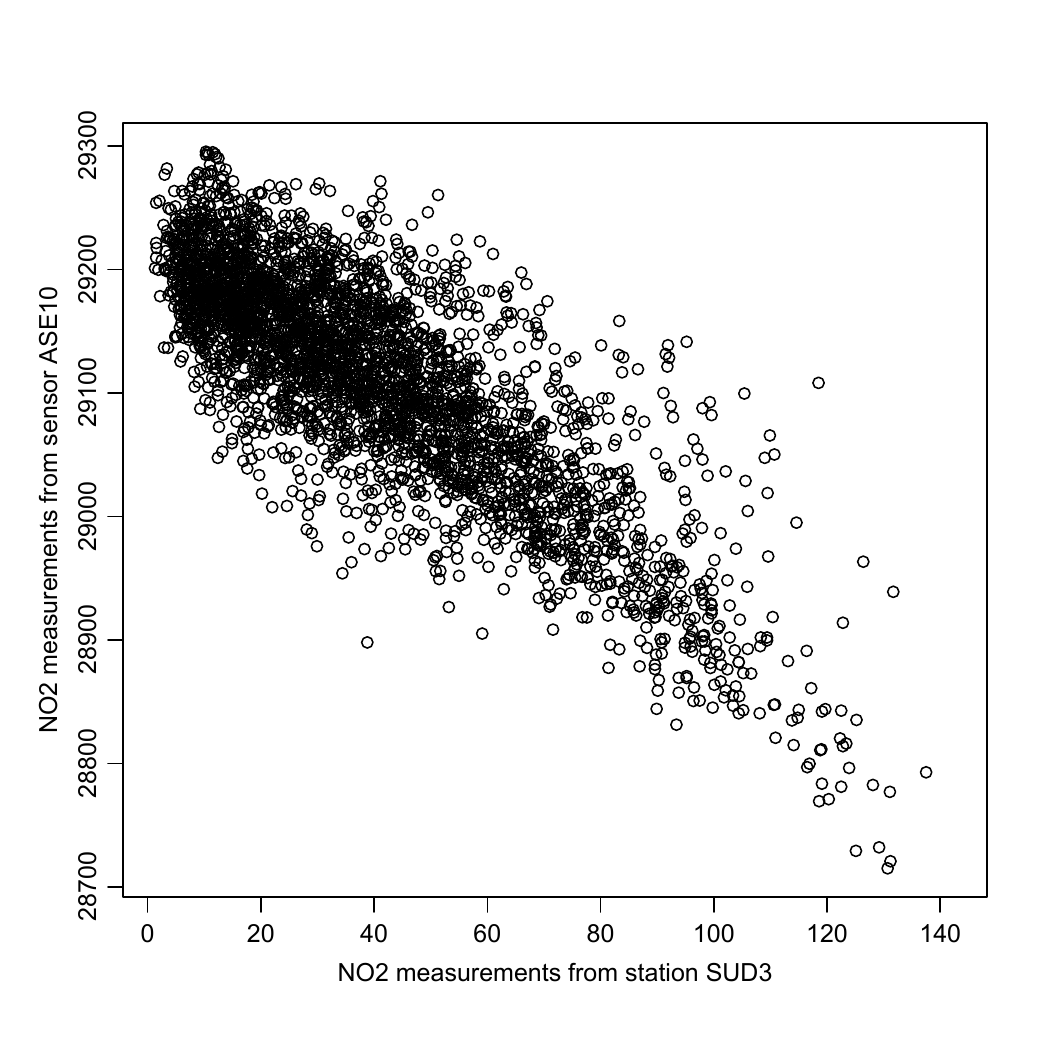}
        \caption{}
        \label{fig:plot_a}
    \end{subfigure}
    \hfill
    \begin{subfigure}[b]{0.48\textwidth}
        \centering
        \includegraphics[width=\textwidth]{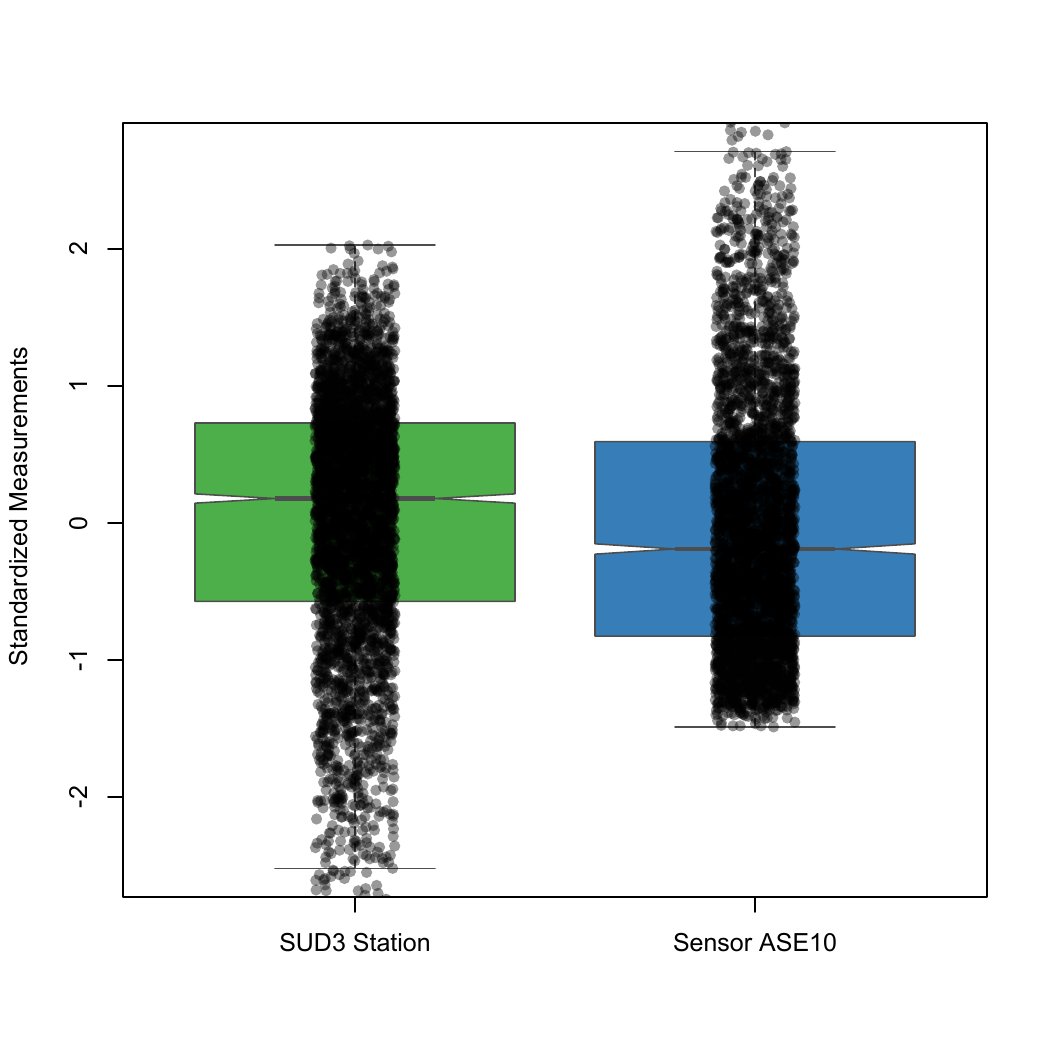}
        \caption{}
        \label{fig:plot_b}
    \end{subfigure}
    \caption{(a) Scatterplot illustrating the relationship between NO$_2$ measurements obtained from the two devices considered in the study. (b) Boxplots used to examine the distributional characteristics of each sample.}
    \label{fig:NO2plots_initial}
\end{figure}

For the purpose of this application, we consider the monitoring station SUD3, which records concentrations in $\mu\mathrm{g}/\mathrm{m}^3$, and the low-cost sensor ASE10, which reports measurements in $\mathrm{mV}$. After removing observations with missing values, the resulting sample size for both variables was 3,739. A preliminary scatterplot is presented in Figure~\ref{fig:NO2plots_initial}(a), where a clear linear association between the two measurement methods is observed. The apparent inverse relationship should be interpreted with caution, as the variables are measured on different scales. Figure~\ref{fig:NO2plots_initial}(b) presents boxplots of the standardized variables, indicating departures from normality in both cases. This observation is confirmed by Fisher’s coefficient of skewness: the SUD3 station exhibits negative skewness ($-0.839$), whereas Sensor ASE10 shows positive skewness ($0.844$).
An important aspect to consider is that the function relating the measurement units of the two devices is unknown and constitutes part of the calibration problem. Consequently, an alternative quantitative approach is required to assess the level of agreement between the SUD3 monitoring station and the ASE10 sensor. We adopt an approach commonly used in environmental sciences, which consists of fitting a regression model between the two variables and subsequently using it to predict, on the SUD3 scale, the values corresponding to measurements obtained from the ASE10 sensor. This approach allows a direct comparison between the measurements obtained from the SUD3 station and the predicted values associated with the ASE10 sensor. The performance of the proposed approach depends on the quality of the fitted model. In the present case, the linear relationship between the variables makes the model fitting a standard procedure from a statistical perspective.

The estimated model using least squares is
\begin{equation}\label{eq:reg_model}
\text{SUD3}=6884.4080 -0.2351 \cdot \text{ASE10}.
\end{equation}
Both regression coefficients are statistically significant at the 5\% level. The global regression (F-test) is highly significant, with a p-value of $2.2\times 10^{-16}$, and the adjusted $R^2$ equals 0.665.

\begin{figure}[htbp]
    \centering
    \begin{subfigure}[b]{0.48\textwidth}
        \centering
        \includegraphics[width=\textwidth]{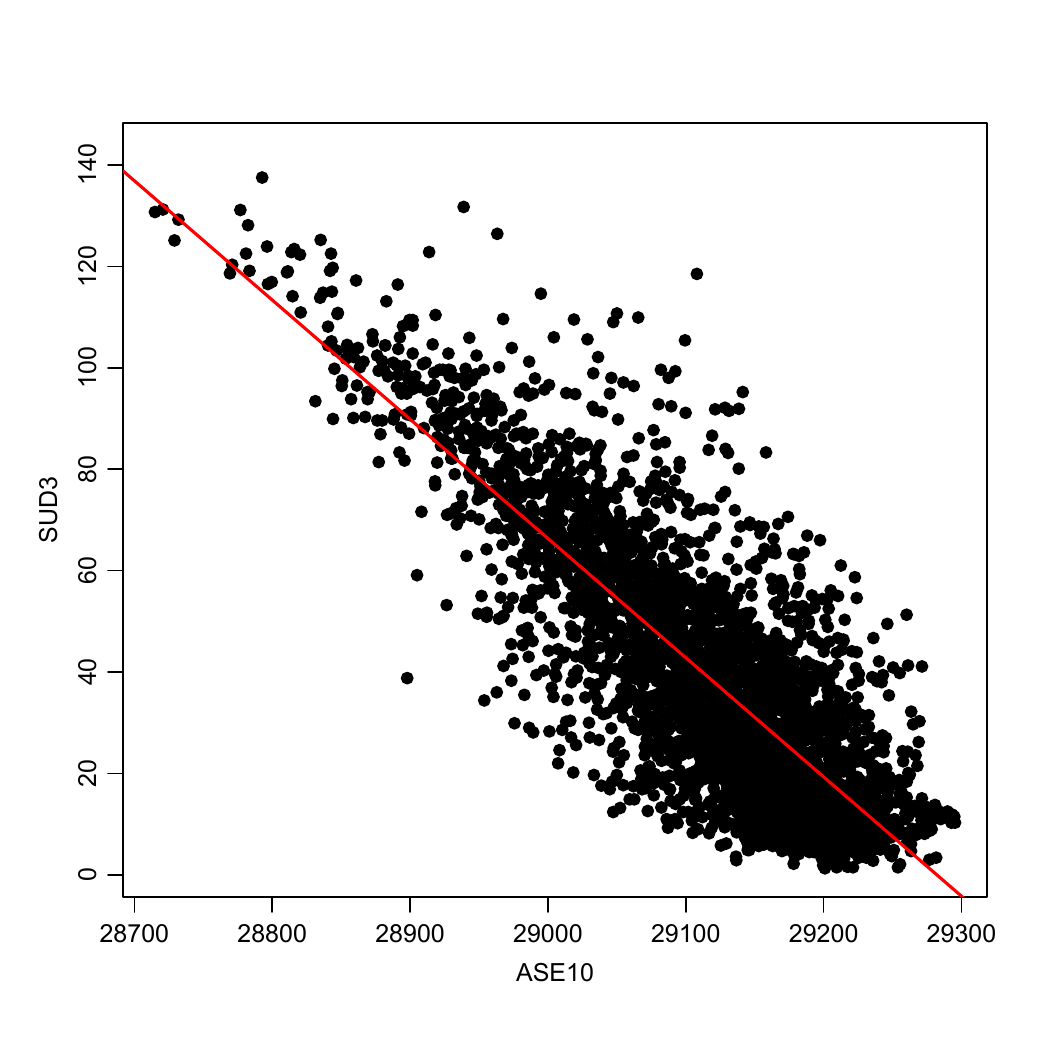}
        \caption{}
        \label{fig:plot_a}
    \end{subfigure}
    \hfill
    \begin{subfigure}[b]{0.48\textwidth}
        \centering
        \includegraphics[width=\textwidth]{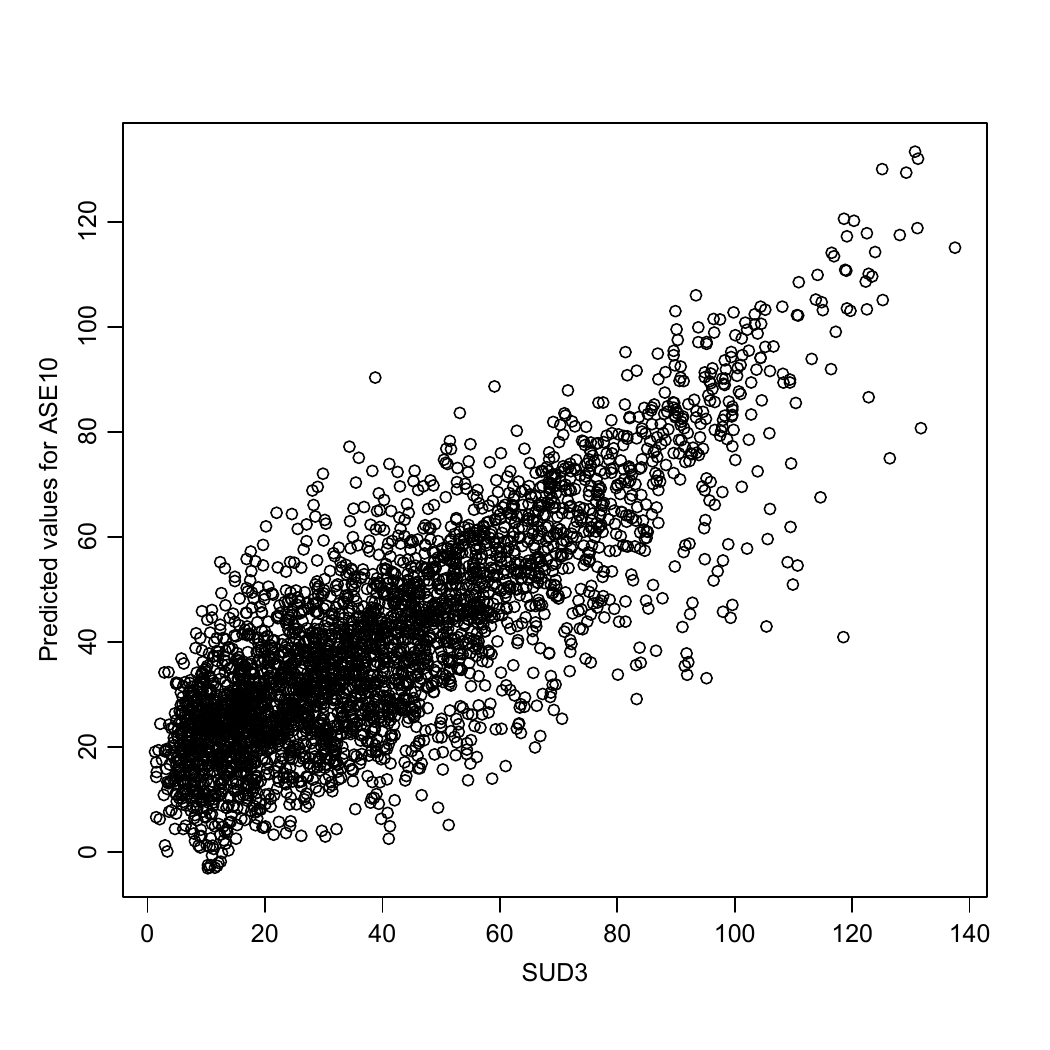}
        \caption{}
        \label{fig:plot_b}
    \end{subfigure}
    \caption{(a) Scatterplot of NO$_2$	
  measurements at station SUD3 versus sensor ASE10, including the fitted regression line.
(b) Scatterplot of observed SUD3 values versus predicted ASE10 values obtained from the regression model \eqref{eq:reg_model}.}
    \label{fig:NO2plots}
\end{figure}
In Figure~\ref{fig:NO2plots}(a), the fitted regression line is displayed together with the observed data. Subsequently, the ASE10 variable is predicted using the linear model in \eqref{eq:reg_model} (see Figure \ref{fig:NO2plots}(b)), and Lin’s concordance correlation coefficient is computed between the observed SUD3 measurements and the corresponding predicted values of ASE10.

The resulting estimate is $\widehat{\rho}_c = 0.798$, with an asymptotic 95\% confidence interval given by $(0.787, 0.809)$. Assessing whether this level of agreement between the sensor and the monitoring station is satisfactory requires expert judgment, taking into account the historical context of the problem, the characteristics and reliability of the devices, and evidence from related studies in the literature.
\begin{figure}[h]
    \centering
    \includegraphics[width=0.5\linewidth]{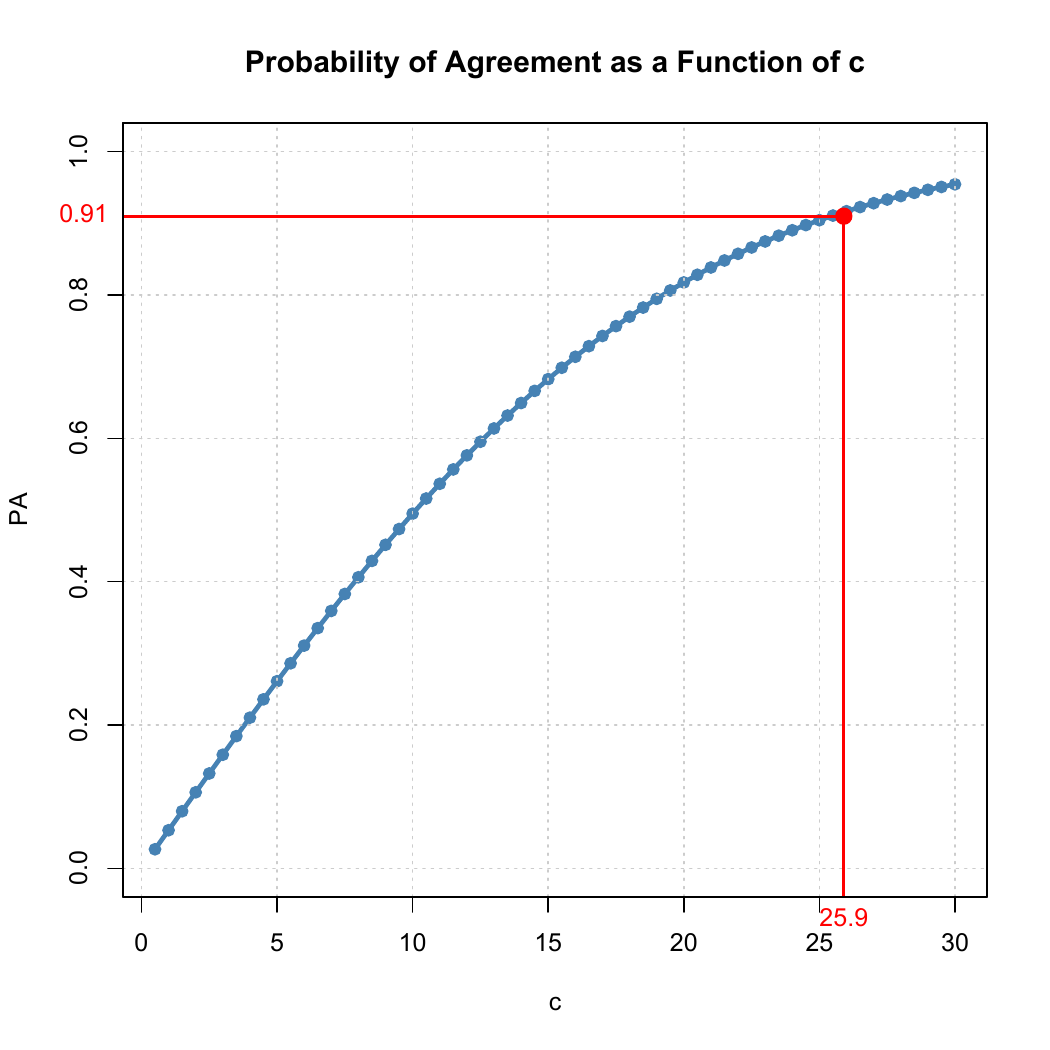}
    \caption{PA between the variables of model  for \ref{eq:reg_model} for different values of $c.$}
    \label{fig:PA_app}
\end{figure}
In addition to Lin’s coefficient, we computed the  PA between the SUD3 variable and the predicted values associated with the ASE10 sensor. Figure~\ref{fig:PA_app} illustrates the behavior of PA as a function of $c$. In this case, the standard deviation of the data from the SUD3 station is 25.9, yielding a corresponding PA value of 0.91. This value is highlighted as a reference choice for $c$. Moreover, both PA and Lin’s coefficient consistently indicate a high level of agreement for this pair of variables.

One aspect that could contribute to improving and refining the analysis is the incorporation of the georeferencing information discussed in Section~\ref{sec:spatial}. For the dataset under consideration, the sensors were deployed at two different locations (the QDP and SUD3 stations), both situated near major roadways to monitor NOx concentrations in Rouen. This configuration makes it challenging to address the problem from a spatial statistics perspective, as there is insufficient information to assess spatial associations among sensors and monitoring stations. Nevertheless, in a pilot study of this nature, it is reasonable for sensors to be placed in close proximity to reference monitoring stations, since the primary objective is calibration. Consequently, positioning sensors far from the stations serving as the measurement gold standard would not be appropriate.

\section{Future Research Directions}\label{sec5}
Methodologies for assessing agreement between two continuous variables have received increasing attention in recent years. However, the rapid emergence of more complex data structures calls for further methodological developments that extend this notion while keeping underlying assumptions to a minimum. Restricting the analysis to purely continuous settings is often insufficient, as many practical problems involve the joint analysis of variables of different types. Extending the concept of agreement to mixed data settings, involving both continuous and discrete variables, therefore remains an open and challenging problem.

One aspect that warrants particular attention is the estimation procedure, especially when dealing with spatial data. Estimation of covariance components can be computationally inefficient, particularly for large datasets. This challenge is often addressed by adopting alternative likelihood-based approaches, such as composite likelihood methods, which have received increasing attention in spatial statistics. When the sample size is large, these and related strategies can substantially reduce computational burden and enable parameter estimation within a reasonable time frame. Although this remains an active area of research, there is still considerable scope for developing more efficient methodologies that, in addition to parameter estimation, provide reliable measures of uncertainty associated with the estimation process.

Another research avenue concerns the assessment of agreement between predictions generated by two neural networks with different configurations that aim to predict the same event. A natural question in this context is the choice of an appropriate distance measure to quantify the discrepancy between the two predictors. This issue can be investigated in the spirit of the agreement-on-the-line framework introduced by \cite{Baek:2022}.

The notion of agreement can be generalized to quantify concordance between two spatial networks. Because the definition of a network is not unique, a natural starting point is to consider two random points in space and then extend the framework to marked point processes \citep{Diggle:2003}. An important aspect that must be addressed is that, when locations are random, the two processes are not necessarily observed at identical spatial coordinates. Consequently, the concept of agreement must be generalized to account for proximity both in terms of spatial location and in terms of characteristics associated with each location. A special application of this framework concerns people’s responses as reflected in the phone calls made during an emergency. If two different types of emergencies are considered, it is natural to seek to link these calling patterns in order to assess whether they differ or are in agreement.


\section*{Acknowledgments}
This work was supported by ANID AC3E CIA250006 and from Fondecyt, grant 1230012.
The author thanks Felipe Osorio, Clemente Ferrer, and Aaron Ellison for their helpful discussions and comments on various topics included in the paper. Their comments have certainly helped to improve the presentation of the manuscript. During the preparation of this work, the authors used AI tools solely to refine the linguistic quality and improve the clarity of the existing text. The AI did not generate any new concepts, data, or original content.

\subsection*{Financial disclosure}

None reported.

\subsection*{Conflict of interest}

The authors declare no potential conflict of interests.

\subsection*{Data Availability Statement}
All data used in this paper are available online.
The original image discussed in Section \ref{motivating} is hosted at \url{https://github.com/JAcostaS/Code-and-Example-Codismap.git}, while the datasets for the application section are available via Mendeley Data at \url{https://data.mendeley.com/datasets/82dnstrd93/1}.

\appendix

\section{The SSIM index}\label{app1}
 Assume that an image can be represented by a matrix $\bm{X}\in \mathcal{M}^{n\times m}(\mathbb{R}^+)$, where $\bm{X}(i,j)$ denotes the gray-level intensity at position $(i,j)$. Alternatively, a realization of $X$ can be described by the vectorization of the matrix $\bm X$, given by $\bm{x} = \mathrm{vec}(\bm{X}) = (\bm{x}^T_1, \ldots, \bm{x}^T_N)^T$, with $N = m \cdot n$ \citep{Brunet:2012a}.

If $\bm{x}, \bm{y} \in \mathbb{R}^N_+$ are two images, the SSIM index is defined as
\begin{equation}
\label{SSIM}
  \text{SSIM}(\bm{x}, \bm{y})=l(\bm{x}, \bm{y})^{\alpha}\cdot c(\bm{x}, \bm{y})^{\beta}\cdot s(\bm{x}, \bm{y})^{\gamma}
\end{equation}
where $\alpha,$ $\beta$ y $\gamma$ are non-negative parameters that are associated with the weight of each multiplicative coefficient

\begin{align}
  l(\bm{x}, \bm{y})&=\left(\frac{2\bar{\bm{x}} \bar{\bm{y}} +c_{1}}{ \bar{\bm{x}} ^{2}+\bar{y}^{2}+c_{1}}\right),\label{eq:lum}\\ 
 c(\bm{x},\bm{y})&=\left(\frac{2 S_{\bm{x}} S_{\bm{y}}+c_{2}}{ {S_{\bm{x}}}^{2}+{S_{\bm{y}}}^{2}+c_{2}}\right),\label{eq:cont} \\
  s(\bm{x},\bm{y})&=\left(\frac{S_{\bm{xy}}+c_{3}}{ S_{\bm{x}}S_{\bm{y}}+c_{3}}\right), \label{eq:cor}
\end{align}
with $\bar{\bm{x}}$, $\bar{\bm{y}}$, $S_{\bm{x}}^2$, $S_{\bm{y}}^2$ and $S_{\bm{xy}}$ denoting the sample means of ${\bm{x}}$ and ${\bm{y}}$, the sample variances of ${\bm{x}}$ and ${\bm{y}}$, and the sample covariance between ${\bm{x}}$ and ${\bm{y}}$, respectively. Commonly, the balanced case is considered, i.e., $\alpha=\beta=\gamma=1$, while acknowledging that these parameters can be estimated from the sample values. The constants $c_1$, $c_2$ and $c_3$ are all positive and can be chosen to preserve the definition of the SSIM index when the denominators are close to zero. Additional mathematical properties of SSIM have been addressed by \cite{Vallejos:2016}, and the estimation of $\alpha$, $\beta$ and $\gamma$ from $\bm{x}$ and $\bm{y}$ can be found in \cite{Osorio:2022}, while a recent application in medical imaging is given in \cite{Maruyama:2023}.

\section{Proof of the Transformation Identities}\label{app2}

\paragraph{Proof of \eqref{eq:trans1}.}
Using a first-order Taylor expansion of $g(\cdot)$ around $\mu_X=\mathbb{E}(X)$, we obtain
$g(X) \approx g(\mu_X) + g'(\mu_X)(X-\mu_X).$ Then 
$\mu_Y = \mathbb{E}[g(X)] \approx g(\mu_X),$
and the variance of $Y$ is
$\sigma_Y^2 = \mathrm{Var}(g(X)) \approx \big(g'(\mu_X)\big)^2 \sigma_X^2.$
Moreover, the covariance between $X$ and $Y$ is
\[
\mathrm{Cov}(X,Y) \approx g'(\mu_X)\,\mathrm{Var}(X)
= g'(\mu_X)\sigma_X^2.
\]
Substituting the above approximations into the definition of $\rho_c$ yields
\[
\rho_c(X,g(X)) \approx
\frac{2\,g'(\mu_X)\,\sigma_X^2}
{\sigma_X^2\left[1+\big(g'(\mu_X)\big)^2\right]
+ \left(\mu_X - g(\mu_X)\right)^2},
\]
which proves \eqref{eq:trans1}. \hfill $\blacksquare$

\paragraph{Proof of \eqref{eq:trans2}.}
Consider a nearly linear transformation of the form
$g(x) = x + \epsilon h(x),$ where $\epsilon \ll 1$ and $h(\cdot)$ is a differentiable function. Then
$g'(\mu_X) = 1 + \epsilon h'(\mu_X),$
$g(\mu_X) = \mu_X + \epsilon h(\mu_X)$, and the numerator of \eqref{eq:trans1} is
$2\,\sigma_X^2\big(1 + \epsilon h'(\mu_X)\big).$
For the denominator, we compute
$1 + \big(g'(\mu_X)\big)^2
= 1 + \big(1 + \epsilon h'(\mu_1)\big)^2
= 2 + 2\epsilon h'(\mu_X) + \epsilon^2 \big(h'(\mu_X)\big)^2,$
and
$\big(\mu_X - g(\mu_X)\big)^2
= \epsilon^2 h(\mu_X)^2.$
Hence, the denominator can be written as
$2\sigma_X^2\left(1 + \epsilon h'(\mu_X)\right)
+ \epsilon^2\!\left[\sigma_X^2 \big(h'(\mu_X)\big)^2 + h(\mu_X)^2\right].$ By Factoring $2\sigma_X^2\left(1 + \epsilon h'(\mu_X)\right)$ from the denominator one obtains
\[
\rho_c(X,g(X)) \approx
\frac{1}
{1 + \frac{\epsilon^2}{2\sigma_X^2}
\left[h(\mu_X)^2 + \sigma_X^2 \big(h'(\mu_X)\big)^2\right]
\left(1 + \epsilon h'(\mu_X)\right)^{-1}}.
\]
Since $\epsilon \ll 1$, we expand $\left(1 + \epsilon h'(\mu_X)\right)^{-1} = 1 + o(1)$ and apply the approximation
$\frac{1}{1+u} = 1 - u + o(u),
 u \to 0.$
This yields
\[
\rho_c(X,g(X)) \approx
1 - \frac{\epsilon^2 h(\mu_X)^2}{2\sigma_X^2}
- \frac{1}{2}\,\epsilon^2 \big(h'(\mu_X)\big)^2
+ o(\epsilon^2).  
\]
\hfill $\blacksquare$

\nocite{*}
\bibliography{wileyNJD-APA}%


\bibliographystyle{plainnat}
\end{document}